\pdfoutput=1
\documentclass[journal]{IEEEtran}

\usepackage[T1]{fontenc}
\ifdefined\pdfgentounicode
  \input{glyphtounicode}
  \pdfglyphtounicode{minus}{002D}
\fi
\usepackage{amsmath}
\usepackage{amsfonts}
\usepackage{graphicx}
\usepackage{tabularx}
\usepackage{cite}
\usepackage{tikz}
\PassOptionsToPackage{hyphens}{url}
\usepackage[hidelinks]{hyperref}

\newcolumntype{P}[1]{>{\raggedright\arraybackslash}p{#1}}
\newcolumntype{L}{>{\raggedright\arraybackslash}X}
\newcolumntype{C}{>{\centering\arraybackslash}X}
\newcolumntype{R}{>{\raggedleft\arraybackslash}X}
\newcommand{\ruledtablestyle}{%
  \footnotesize
  \setlength{\tabcolsep}{3pt}%
  \renewcommand{\arraystretch}{1.25}%
}
\newcommand{\stackedtablecell}[1]{%
  \begingroup
  \setlength{\tabcolsep}{0pt}%
  \begin{tabular}[c]{c}#1\end{tabular}%
  \endgroup
}
\newcommand{\intervalentry}[2]{\stackedtablecell{$#1$\\{\scriptsize$[#2]$}}}
\newcommand{\refdoi}[1]{doi: \href{https://doi.org/#1}{\nolinkurl{#1}}}
\newcommand{\refonline}[1]{[Online]. Available: \url{#1}}
\newtheorem{lemma}{Lemma}
\newtheorem{proposition}{Proposition}
\newcounter{example}

\begin{document}
\title{\fontsize{23bp}{28bp}\selectfont
When Replay Misranks Vulnerability Patching Rules:\\
A Game-Theoretic Analysis of Attacker Response}

\author{\IEEEauthorblockN{Yuning Jiang\IEEEauthorrefmark{1},
Nay Oo\IEEEauthorrefmark{2},
Qiaoran Meng\IEEEauthorrefmark{1},
Lu Lin\IEEEauthorrefmark{1},
Dusit Niyato\IEEEauthorrefmark{3},
Zehui Xiong\IEEEauthorrefmark{4},
Hoon Wei Lim\IEEEauthorrefmark{2},
Biplab Sikdar\IEEEauthorrefmark{1}
}

\IEEEauthorblockA{\IEEEauthorrefmark{1}
National University of Singapore, Singapore}
\IEEEauthorblockA{\IEEEauthorrefmark{2}NCS Cyber Special Ops R\&D, Singapore}
\IEEEauthorblockA{\IEEEauthorrefmark{3}Nanyang Technological University, Singapore}
\IEEEauthorblockA{\IEEEauthorrefmark{4}Queen's University Belfast, Belfast, UK}
}

\maketitle

\begin{abstract}
Fixed-response replay scores vulnerability patching rules against exploitation recorded under the status quo, holding attacker behavior fixed. Yet reallocating a limited patching budget changes how long vulnerabilities stay exposed, and hence which exploits are worth developing. We study when this response reverses replay's ranking in a synthetic leader--follower game: defenders commit to a budgeted rule, an attacker observes it and develops exploits at a cost, and observed exploitation triggers delayed catalog listing. Policy-specific evaluation instead lets the attacker respond to each candidate rule. Under full adoption, replay and policy-specific evaluation correspond to believed adoption shares of zero and one. Prespecified tests on fresh instances show that the lower-loss rule depends on development cost and publication-score quality. At an intermediate cost, replay favors repairing listed vulnerabilities first (evidence-first) and policy-specific evaluation favors ranking by publication score (score-first); a descriptive sweep places the change at 43\% believed adoption (95\% interval 32\% to 54\%). At a second intermediate cost, replay favors ranking by a frozen exploitation predictor (probability-only), but under the attacker's response the predictor's concordance falls and the rule's loss exceeds score-first's by 11.9\% of status-quo loss. In descriptive extensions, refitting on each latest round makes this loss alternate. Pooling or damping curbs the alternation; even so, after ten refits probability-only ranking beats score-first only where high cost makes development sparse. With two attribute shares from public vulnerability records, replay still credits it with gains under sparse development that are not established after adoption. Costs and score quality are synthetic and uncalibrated.
\end{abstract}


\IEEEpeerreviewmaketitle

\section{Introduction}\label{sec-1}

Security teams cannot patch every disclosed vulnerability at once, so they must rank them, and the public signals they rank by arrive at different times. The Known Exploited Vulnerabilities (KEV) Catalog of the U.S. Cybersecurity and Infrastructure Security Agency (CISA) lists a vulnerability after exploitation has been observed~\cite{ref-1}. The Common Vulnerability Scoring System (CVSS) summarizes severity~\cite{ref-2} and the Exploit Prediction Scoring System (EPSS) forecasts observed exploitation activity~\cite{ref-3}, so both can guide a decision before such evidence exists. How much weight each signal should carry is an open policy question. Binding Operational Directive (BOD) 22-01 set deadlines for U.S. federal civilian agencies to remediate cataloged vulnerabilities~\cite{ref-27}. Its 2026 replacement, BOD 26-04, keeps KEV status within a broader framework that also uses public exposure, exploit automatability, and technical impact~\cite{ref-28}. The economics of exploitation may be shifting as well. OpenAI reports that an automated Linux kernel audit in a controlled setting produced working local privilege-escalation exploits, averaging \$359 in API credits per successful exploit for the development stage and \$1,279 for the full workflow, figures it presents as upper bounds~\cite{ref-5}. If automation lowers development costs, attacks that were previously uneconomical could become worthwhile.

Patching rules are commonly compared by replaying them on a record of past exploitation and asking which would have left less exposure. A careful replay is time-aligned, so each rule uses only the information available on the decision date. It still holds the attacker's behavior fixed, because both rules face the events generated under the status quo. That assumption is doubtful once a rule is widely adopted. The rule changes how long each vulnerability stays exposed, which can change which exploits are worth developing. Work on performative prediction studies this feedback for forecasts, whose deployment can change the events they predict~\cite{ref-4}. We ask the corresponding question for patching rules. \textbf{When can attacker response reverse the ranking that fixed-response replay gives two patching rules?}

We study the question in a synthetic leader--follower game. A population of defenders shares a public catalog. Vulnerabilities arrive with publication scores, and in each period defenders divide a limited patch budget between listed vulnerabilities and a score-ordered queue. The attacker observes the rule and chooses which exploits to develop, weighing each vulnerability's development cost against the returns from the exposure the rule leaves. Exploitation in the model means that development is completed within the evaluation horizon. Observed exploitation leads to listing after a delay, and a listing redirects later budgets. We vary the development cost, the quality of the publication score as a ranking of loss weights, and the share of defenders that adopt a rule.

Each pair of rules is evaluated in two ways on the same synthetic instances and budgets. In \textbf{fixed-response replay}, both rules face the attacker plans generated under a status-quo policy. In \textbf{policy-specific evaluation}, each rule faces the plans induced by its own adoption. A reversal occurs when the rule with lower expected loss in replay has higher expected loss under the policy-specific responses. Attacker plans are enumerated exactly, an independent period-level implementation checks small instances, and the comparisons behind our main claims were specified before they were run on fresh instances.

We report three experimental findings and one theoretical characterization. Magnitudes are shares of status-quo loss, the expected loss when defenders keep the status-quo policy and the attacker responds to it.

\begin{enumerate}

\item \textbf{The lower-loss rule depends on development cost and score quality (Section~\ref{sec-4}).} With an exact publication score, ranking by score (score-first) has lower loss than patching listed vulnerabilities first (evidence-first) at the low costs tested, and evidence-first has lower loss at the high costs. At a fixed cost, a noisier score reverses the comparison.

\item \textbf{Fixed-response replay can reverse the order of two rules (Section~\ref{sec-5-1}).} At an intermediate cost, replay favors evidence-first by 3.0\% of status-quo loss, whereas policy-specific evaluation favors score-first by 4.7\%. The two evaluations agree at a lower and at a higher tested cost. Descriptive analyses attribute most of the swing to how many exploits the attacker develops and place the change of ranking at a believed adoption share of about 43\% (95\% bootstrap interval 32\% to 54\%).

\item \textbf{Forecast accuracy and patching loss can move apart (Section~\ref{sec-5-2}).} For an exploitation predictor fitted under the status quo and then frozen, replay at an intermediate cost favors ranking by predicted probability alone (probability-only) over score-first by 2.6\% of status-quo loss. Under the attacker's response to that rule, the predictor's concordance is 0.568, down from 0.653 under the response to score-first, and the rule's loss exceeds score-first's by 11.9\% of status-quo loss. Weighting the predicted probability by the publication score retains a gain of 1.1\% after adoption. In descriptive refits over ten rounds, using the latest round alone makes the loss alternate; pooling or damping curbs the alternation, yet the refitted rule beats score-first only where high cost makes development sparse.

\item \textbf{Believed adoption links the two evaluations (Section~\ref{sec-3-5}).} Under full adoption, replay and policy-specific evaluation are the two ends of one family indexed by the adoption share the attacker believes. Under exact best response, any drop from replay loss to policy-specific loss is bounded by the attacker's savings in development cost, so replay cannot overstate a rule's loss when development is costless. With finite response sensitivity, opposite rankings at the two ends imply at least one change of ranking in between. When few defenders adopt, replay converges for the adopters' own loss while the population-wide ranking can still differ.

\end{enumerate}

These results show that replay reversals occur under specified conditions in the model. They do not measure how often reversals occur in practice, which would require evidence on how attackers respond to deployed patching policies. Development costs and score quality are synthetic and are not calibrated to data, the publication score is on a loss-weight scale, and the predictor is trained on synthetic development probabilities. In a descriptive variant that takes two attribute shares from public vulnerability records, opposite rankings were supported at costs with dense development, while replay misstated the size of the loss gap at costs with sparse development (Section~\ref{sec-6-2}).

\section{Related work}\label{sec-2}

\subsection{Vulnerability prioritization and time-aligned evaluation}\label{sec-2-1}

Empirical studies of prioritization rank vulnerabilities against observed exploitation, and they show how much the answer depends on the date at which a rule is allowed to see its inputs. Allodi and Massacci used case-control comparisons to ask which vulnerability attributes distinguish vulnerabilities exploited in the wild. Adding information on the availability of proof-of-concept code or on exploits appearing in black markets improved prioritization over high CVSS severity alone~\cite{ref-7}.

Jacobs et al. compared remediation heuristics and learned predictors in terms of how many vulnerabilities they select and what share of vulnerabilities with observed exploitation that set covers~\cite{ref-8}. The original EPSS paper estimated the probability of exploitation in the wild within 12 months of public disclosure~\cite{ref-epss-original}. The later EPSS evaluation included KEV membership among six simple heuristics and compared CVSS v3.x base scores with EPSS versions 1 to 3 at matched remediation effort and matched coverage, against exploitation activity observed in the following 30 days~\cite{ref-9}.

Suciu et al. distinguish the development of functional exploits from their use in the wild. Their Expected Exploitability score uses dated evidence, and their evaluation compares rankings at disclosure and later dates to measure the value of information as it arrives~\cite{ref-10}. These studies establish the importance of specifying both the decision date and the outcome against which a ranking is evaluated. In their comparisons, the observed exploit outcomes provide a common reference for evaluating alternative rankings. We ask what happens to that reference when the rule under evaluation is widely adopted.

\subsection{Strategic models of patching and defense}\label{sec-2-2}

Game-theoretic models make the attacker's choice depend on where the defender allocates protection, which is the dependence that replay holds fixed. Gianini et al. cast constrained vulnerability patching as a search game~\cite{ref-11}. In their two-vulnerability simultaneous game, the defender chooses one vulnerability to patch and the attacker chooses one to attack. They derive a mixed-strategy Nash equilibrium for the case with no pure-strategy equilibrium, extend the simultaneous formulation, and present a two-round example with alternating moves and partial observation. The choices in those examples are individual patch and attack actions. Our comparison uses budgeted rules that act on public scores and a catalog whose entries follow development in the game.

Stackelberg security games represent a defender's commitment to a resource allocation and an adversary's response to that commitment. Sinha et al. survey models of this interaction, along with solution methods and deployed applications~\cite{ref-sinha-survey}. Nguyen et al. compare strategies based on quantal-response models with robust alternatives in human-subject experiments~\cite{ref-12}. Haghtalab et al. give conditions under which response observations from sufficiently distinct defender allocations can support learning a parametric response model~\cite{ref-13}. Our game uses a specified logit response over development plans to evaluate changes in rule order. Its response distributions come from the model's payoffs.

An et al. study limited surveillance in security games. Their attacker updates a belief about the defender's strategy after observing defender actions~\cite{ref-an-surveillance}. In Section~\ref{sec-3-5}, we take the attacker's belief about the share of organizations adopting a rule as given and analyze how that belief affects rule rankings.

The timing and allocation of security effort also appear in economic models. Arora, Telang, and Xu analyze how a disclosure deadline affects a vendor's choice of when to release a patch~\cite{ref-14}. Cavusoglu, Cavusoglu, and Zhang model patch release by a vendor and installation by a firm, and study when cost sharing can coordinate those decisions~\cite{ref-cavusoglu-patch}. Allodi, Massacci, and Williams model how fixed exploit development costs can delay development and limit the vulnerabilities mass attackers target. They test the model's predictions on observed attack signatures~\cite{ref-work-averse}. Bono models an attacker and a defender investing in AI capabilities, including how one organization's defense can redirect the attacker to another~\cite{ref-15}. Our experiment varies a development-cost parameter and compares remediation rules within a fixed budget.

\subsection{Feedback from deploying a decision rule}\label{sec-2-3}

A third literature treats the data used for evaluation as a consequence of the deployed rule. Lucas argued that behavioral relationships estimated under one policy can change when the policy changes, limiting their use for counterfactual evaluation~\cite{ref-lucas}. Hardt et al. model how people change inputs at a cost after learning a classifier's rule~\cite{ref-16}. Perdomo et al. formalize how deployment can change a data distribution and distinguish evaluation on a distribution induced by some reference decision from evaluation on the distribution induced by the decision being scored. The former is a fixed slice of their decoupled risk. The latter is its diagonal, the performative risk~\cite{ref-4}.

Our comparison instantiates that distinction for budgeted patching. Fixed-response replay is the slice at the status-quo response, and policy-specific evaluation is the diagonal. We ask when the two rank a pair of rules differently, with the response specified by an attacker model and dated exploitation evidence feeding back into remediation.

\section{Model}\label{sec-3}

Figure~\ref{fig:model-schematic} shows how a development plan determines when exploitation and catalog listing begin and how each evaluation chooses which plans to use.

\begin{figure*}[!t]
\centering
\begin{tikzpicture}[
    font=\footnotesize,
    x=1cm, y=1cm,
    event/.style={align=center, inner sep=1pt, anchor=north},
    route/.style={align=center, inner sep=2pt, minimum height=9mm,
        rounded corners=2pt, draw=black!45},
]
\definecolor{schematicblue}{RGB}{53,80,97}
\definecolor{schematicfill}{RGB}{235,232,222}

\node[anchor=west, font=\footnotesize\bfseries] at (0,0)
    {One vulnerability under one development plan};
\draw[schematicblue, thick, ->] (0.1,-1.03) -- (8.55,-1.03);
\foreach \position/\time in {0.65/{\tau_i},2.67/{w_i},4.73/{w_i+1},7.14/{\lambda_i=w_i+1+L}} {
    \fill[schematicblue] (\position,-1.03) circle[radius=1.5pt];
    \node[above=3pt, text=schematicblue] at (\position,-1.03) {$\time$};
}
\node[event] at (0.65,-1.18) {Publication};
\node[event] at (2.67,-1.18) {Development ends\\Exploitation starts};
\node[event] at (4.73,-1.18) {Evidence\\recorded};
\node[event] at (7.14,-1.18) {Catalog listing\\(if a remedy exists\\and $\lambda_i<T$)};
\draw[schematicblue, thick, ->] (7.14,-2.25) -- (7.14,-2.60);
\node[route, fill=schematicfill, minimum width=8.35cm] at (4.32,-3.12)
    {Listed vulnerabilities draw share $\theta$ of the patch budget.\\
     The balance goes to the score-ordered queue.};

\draw[black!30] (8.88,0.16) -- (8.88,-3.57);
\begin{scope}[xshift=9.18cm]
\node[anchor=west, font=\footnotesize\bfseries] at (0,0)
    {Two ways to evaluate candidate rules at full deployment};
\node[anchor=west, font=\footnotesize\bfseries] at (0,-0.58)
    {Fixed-response replay};
\node[route, minimum width=4.35cm] (replay) at (2.24,-1.28)
    {Both rules draw plans from the\\status-quo response ($a=0$).};
\node[route, fill=schematicfill, minimum width=3.65cm] (replayloss) at (6.68,-1.28)
    {Recompute loss under\\each deployed rule.};
\draw[schematicblue, thick, ->] (replay) -- (replayloss);

\node[anchor=west, font=\footnotesize\bfseries] at (0,-2.13)
    {Policy-specific evaluation};
\node[route, minimum width=4.35cm] (own) at (2.24,-2.83)
    {Each rule draws plans from the\\response it induces ($a=1$).};
\node[route, fill=schematicfill, minimum width=3.65cm] (ownloss) at (6.68,-2.83)
    {Recompute loss under\\each deployed rule.};
\draw[schematicblue, thick, ->] (own) -- (ownloss);
\end{scope}
\end{tikzpicture}
\caption{Plan timing and the two evaluations at full actual adoption. The attacker believes the adoption share is $a$.}
\label{fig:model-schematic}
\end{figure*}

\subsection{Vulnerability arrivals and catalog listing}\label{sec-3-1}

The game runs for eight discrete periods, $t=0,\ldots,T-1$, with $T=8$. Each main experiment instance has $n=8$ synthetic vulnerabilities. Vulnerability $i$ is published at $\tau_i$ and has $N_i$ exposed vulnerability--host instances. Its loss per exposed instance per period is $s_i=\rho_i(1+z_i)$, where $\rho_i$ is the reachable fraction and $z_i$ indicates whether exploitation can be automated. The instance also assigns a development difficulty $c_i$ and an indicator $m_i$ for whether a remedy exists. Table~\ref{tab:notation} lists the symbols used most often.

\begin{table}[!t]
\caption{Symbols used most often.}\label{tab:notation}
\centering
\ruledtablestyle
\begin{tabularx}{\columnwidth}{|l|L|}
\hline
\textbf{Symbol} & \textbf{Meaning} \\ \hline
$n$, $T$ & Vulnerabilities per instance and number of periods \\ \hline
$\tau_i$, $N_i$ & Publication period and exposed instances of vulnerability $i$ \\ \hline
$s_i$, $\hat s_i$ & Loss weight per exposed instance per period, and its publication score \\ \hline
$q$ & Score quality: probability that the scores order two unequal loss weights correctly \\ \hline
$\pi$, $\pi_0$ & Candidate patching rule and status-quo rule \\ \hline
$\theta$ & Share of the patch budget reserved for listed vulnerabilities \\ \hline
$\alpha$, $a$ & Actual adoption share, and the share the attacker believes \\ \hline
$L$, $K$ & Listing delay in periods, and development completions allowed per period \\ \hline
$S$, $C(S)$ & Attacker plan (the vulnerabilities developed) and its development cost \\ \hline
$\kappa$ & Relative development cost $F/F_0$ \\ \hline
$\beta$ & Sensitivity of the logit response; $\beta=\infty$ is exact best response \\ \hline
$\ell_\pi(S)$ & Loss under rule $\pi$ when the attacker follows plan $S$ \\ \hline
$J(\pi\mid r)$ & Expected loss of $\pi$ when plans follow the response to rule $r$ \\ \hline
$D$, $G$ & Evidence-first minus score-first loss, and a rule's gain over score-first \\ \hline
\end{tabularx}
\end{table}

We generate publication periods uniformly from $\{0,1,2,3\}$ and exposure counts uniformly from $\{20,\ldots,100\}$. The probability of internet exposure is $0.3$, with $\rho_i=1$ for internet-facing vulnerabilities and $0.25$ otherwise. The probabilities of automation and a remedy are $0.3$ and $0.8$, respectively. Development difficulty is uniform on $[0.5,1.5]$. These distributions and cost scales are synthetic.

Let $w_i$ be the completion period for a vulnerability the attacker develops. Exploitation starts in period $w_i$, and the model records evidence one period later. A vulnerability with a remedy enters the public catalog $L$ periods after evidence is recorded, at $\lambda_i=w_i+1+L$, provided that $\lambda_i<T$. Each plan thus determines its own catalog calendar through its development schedule. Development completion starts the evidence clock even if exposure has already fallen to zero. This can happen only with a finite development limit. With $K=\infty$, a selected vulnerability completes development at publication ($w_i=\tau_i$). Loss is assessed before repairs in that period, so $U_i(w_i)=N_i>0$ when the clock starts. Among the main experiments, this case is possible in the $K=1$ half of the cost and score-quality contrasts and of the policy-selection experiment. The registered replay reversals and the believed-adoption, refit, public-record, and larger-instance analyses all use $K=\infty$.

At the start of period $t$, exploitation incurs loss $s_i U_i(t)$ for each vulnerability with $w_i\leq t$, where $U_i(t)$ is the number of unpatched instances across both groups before that period's repairs. Defenders then patch, using catalog entries available through period $t$. For example, when $L=1$, completion at $w_i$ produces evidence at $w_i+1$ and a listing at $w_i+2$ if the vulnerability is eligible. Patching in the listing period first affects the loss calculation in the following period.

\subsection{Defender population and patching rules}\label{sec-3-2}

For each vulnerability, a fraction $\alpha$ of exposed instances belongs to organizations that adopt the candidate rule. The rest belong to organizations that retain the status-quo rule $\pi_0$. Both groups see the same catalog. The total budget each period is $B=0.05\sum_i N_i$, measured in vulnerability--host instances. The groups use separate budgets of $\alpha B$ and $(1-\alpha)B$. Before spending these budgets, each group repairs $5\%$ of its remaining exposed instances for each published vulnerability. This baseline repair leaves its budget intact. Exposure and repair amounts evolve as continuous quantities.

The two-layer rule allocates a fraction $\theta$ of the budget to listed vulnerabilities. It repairs the same fraction of each listed vulnerability's remaining instances, up to full coverage. The rest of the budget, including any unused evidence allocation, goes to published vulnerabilities in descending score order. The group exhausts each vulnerability's remaining instances before moving to the next, breaking score ties by index. Any budget left at the end of the period expires. Table~\ref{tab-1} summarizes the rules.

\begin{table}[!t]
\caption{Patching rules for the budget after baseline repair.}\label{tab-1}
\centering
\ruledtablestyle
\begin{tabularx}{\columnwidth}{|P{.31\columnwidth}|L|}
\hline
\textbf{Rule} & \textbf{Use of the budget after baseline repair} \\ \hline
Evidence-first ($\pi_E$, $\theta=1$) & Repair listed vulnerabilities proportionally to their remaining counts, then spend any balance in publication-score order. \\ \hline
Score-first ($\pi_S$, $\theta=0$) & Spend the full budget in publication-score order. \\ \hline
Evidence-plus-score ($\pi_{ES}$) & Repair listed vulnerabilities in publication-score order, then repair other published vulnerabilities in that same order. \\ \hline
Status quo ($\pi_0$, $\theta_0=0.8$) & Allocate $80\%$ to the proportional evidence layer and the rest, including any unused evidence allocation, to the publication-score queue. \\ \hline
\end{tabularx}
\end{table}

The publication score is $\hat s_i=s_i\exp(\sigma\varepsilon_i)$, with independent standard normal noise $\varepsilon_i$. Score quality $q$ is the probability that the publication scores correctly order a randomly drawn pair of vulnerabilities from the model population, conditional on the pair having unequal loss weights. An exact score ($q=1$) uses $\sigma=0$. We reuse each instance's noise draws across quality levels and vary only $\sigma$. Both groups and the attacker see the same publication scores. The scores affect repair priority, while losses and development scheduling use the true weights.

In the prediction experiment, adopters use a score queue alone. We compare descending order by $\hat s_i$, by a frozen prediction $\hat p_i$ of exploitation, and by their product $\hat p_i\hat s_i$~\cite{ref-artifact}. The attacker observes the rule and the predictions. The nonadopting group retains the status-quo policy with its original publication-score queue.

\subsection{Attacker response and defender loss}\label{sec-3-3}

The game has a defender who commits to a repair rule for the adopting group and an attacker who then chooses a development plan. The defender's payoff is minus the loss defined below. We assume that the attacker knows the synthetic instance, the rules of the game, the defender's commitment, and the actual adoption share before choosing a plan. Section~\ref{sec-3-5} separates believed adoption from actual adoption.

An attacker plan is a subset of vulnerabilities. Each plan has a fixed development schedule given the instance and the development limit $K$. The schedule respects release dates and allows at most $K$ completions per period. It processes vulnerabilities by publication period, then by descending $s_iN_i$ and by index, assigning each to the earliest period with room. With $K=\infty$, development for every selected vulnerability finishes at publication. A selected vulnerability whose development falls outside the horizon generates neither exploitation nor development cost.

Given a rule and a plan, defender loss $\ell_\pi(S)$ is the sum of period losses across both groups. One loss point is the modeled loss from one unpatched vulnerability--host instance with weight $s_i=1$ in one period of exploitation. The attacker's net payoff is this gross return minus the one-time cost $F c_i v_{\mathrm{unit}}$ for each completed development, where $v_{\mathrm{unit}}=n^{-1}\sum_i s_iN_i(T-\tau_i)$. This scale is the mean loss per vulnerability if all development finishes at publication and exposure stays unchanged. We write relative cost as $\kappa=F/F_0$. The anchor $F_0=2^{1/4}$ is the point of the cost grid $\{2^{k/4}\}$ at which an exact best response to the status quo develops closest to 10\% of vulnerabilities on 50 calibration instances. It was fixed before the experiments and has no empirical counterpart~\cite{ref-artifact}. Development costs enter only the attacker's payoff.

We enumerate all $2^n$ plans. For finite $\beta$, the probability of plan $S$ under rule $\pi$ is proportional to $\exp(\beta u_\pi(S)/v_{\mathrm{unit}})$, where $u_\pi(S)$ is its net payoff. Thus $\beta=0$ gives uniform plan probabilities, and $\beta=\infty$ selects a payoff-maximizing plan, with ties going to the lowest plan index~\cite{ref-artifact}. A change in patching priorities can change the loss and net payoff of the same plan, and hence its probability.

\subsection{Evaluation methods}\label{sec-3-4}

\subsubsection{Fixed-response replay and policy-specific evaluation}\label{sec-3-4-1}

For a given synthetic instance and configuration, let $P_r(S)$ be the probability of plan $S$ when the attacker responds to rule $r$. Let $\ell_\pi(S)$ be the loss from evaluating rule $\pi$ on that plan with its own development dates and the resulting listing calendar. We use the same instance and budget when changing either rule. Write the expected loss as

\[
J(\pi\mid r) = \sum_S P_r(S)\,\ell_\pi(S).
\]

\begin{table*}[!t]
\caption{Response distributions and recomputed quantities in the two evaluation methods.}\label{tab-2}
\centering
\ruledtablestyle
\begin{tabularx}{\textwidth}{|P{.23\textwidth}|L|L|}
\hline
\textbf{Evaluation} & \textbf{Attacker plans and catalog} & \textbf{Repairs and losses} \\ \hline
\textbf{Fixed-response replay}, $J(\pi\mid\pi_0)$ & Use the distribution calculated under the status quo $\pi_0$ for both candidate rules. Each plan carries its development schedule, evidence dates, and listing calendar. & Run each rule on the common plans and calendars with the same patching budget. Recompute repairs, remaining exposure, and losses under that rule. \\ \hline
\textbf{Policy-specific evaluation}, $J(\pi\mid\pi)$ & Recompute the attacker's plan payoffs and response distribution for each candidate rule. Each realized plan still carries its own schedule and calendar. & Apply the candidate rule to its adopting group and the status quo to the rest. Average the losses using the response probabilities for that candidate. \\ \hline
\end{tabularx}
\end{table*}

Figure~\ref{fig:model-schematic} shows where the plans come from at full adoption. With endogenous listing, each plan generates its own calendar in both evaluations in Table~\ref{tab-2}. For a given plan, its completion and listing dates stay fixed when we change the rule being evaluated. Policy-specific evaluation changes the probability of the plan itself. This can change which vulnerabilities are developed, how many development completions occur, and which dates and catalog entries the defenders encounter. In partial adoption, a different shared catalog can also change the repairs and losses of the group that retains the status quo. The two evaluations are counterfactuals within the synthetic game, with the replay record generated by the model.

For evidence-first versus score-first, we use the signed loss differences

\[
\begin{aligned}
D_{\mathrm{replay}} &= J(\pi_E\mid\pi_0)-J(\pi_S\mid\pi_0),\\
D_{\mathrm{adopt}} &= J(\pi_E\mid\pi_E)-J(\pi_S\mid\pi_S).
\end{aligned}
\]

Positive differences favor score-first. Negative differences favor evidence-first. The two evaluations reverse the ranking when the signed differences have opposite signs. Their difference, $D_{\mathrm{adopt}}-D_{\mathrm{replay}}$, measures the change in the rule comparison when each rule faces its own response, relative to the status-quo reference. It combines any changes in which vulnerabilities are developed, their number, and the timing of development and listing. Replaying under a different common response changes the reference, so we report those comparisons separately as descriptive checks.

\subsubsection{Evaluation of a frozen exploitation predictor}\label{sec-3-4-2}

For a given rule, the exploitation probability $p_{i,r}$ is the sum of $P_r(S)$ over plans in which development for vulnerability $i$ finishes within the horizon. We fit one logistic predictor per synthetic environment, using exact probabilities under the status quo as labels on a separate fitting set. The features available at publication are release period, log exposure count, log publication score, remedy availability, and log public exposure value relative to the mean among vulnerabilities published by that period. Public exposure value is $\hat s_iN_i(T-\tau_i)$. We standardize features using fitting-set moments and minimize mean cross-entropy with an $L_2$ penalty of $10^{-6}/2$ times the squared norm of the coefficients, including the intercept. We freeze the fitted predictor for evaluation under every compared rule in that environment.

We compare the same predictions with the exact probabilities induced by each rule. Within an instance, concordance is the fraction of pairs with unequal induced probabilities that the predictor orders correctly, with half credit for tied predictions. For the paired change in concordance, an instance contributes zero when either response has no comparable pairs. We report the number of such instances and compute stand-alone concordance means over instances with comparable pairs. Concordance measures ranking by exploitation probability. The publication-score quality $q$ measures ranking by per-instance loss weight.

For the prediction experiment, let $\pi$ be the probability-only or product rule. We express the gain relative to score-first as

\[
\begin{aligned}
G_{\mathrm{replay}}(\pi) &= J(\pi_S\mid\pi_0)-J(\pi\mid\pi_0),\\
G_{\mathrm{adopt}}(\pi) &= J(\pi_S\mid\pi_S)-J(\pi\mid\pi).
\end{aligned}
\]

Positive gains favor the tested rule. We also replay both rules under the score-first response. Relative to that replay, policy-specific evaluation changes only the response faced by the tested rule. Concordance and patching loss are separate outcomes.

\subsubsection{Prespecified comparisons and uncertainty}\label{sec-3-4-3}

\begin{table*}[!t]
\caption{Parameters and prespecified tests for the rule comparisons.}\label{tab-3}
\centering
\ruledtablestyle
\begin{tabularx}{\textwidth}{|P{.15\textwidth}|c|c|c|c|c|c|P{.21\textwidth}|L|}
\hline
\textbf{Experiment} & \stackedtablecell{\textbf{Instances}\\\textbf{(fit/evaluation)}} &
\textbf{$q$} & \textbf{$\alpha$} & \textbf{$K$} & \textbf{$\beta$} & \textbf{$L$} &
\textbf{Tested costs ($\kappa$)} & \textbf{What is tested} \\ \hline
Cost and score quality & -- / 400 & 1.00, 0.55 & 1 & $1,\infty$ & 4 & 1 &
0.250, 0.354, 0.500 (low)\newline 1.414, 2.000 (high)\newline 0.707 (fixed) &
Rule rankings at low and high costs for $q=1.00$ and at the fixed cost for both qualities. \\ \hline
Reversal brackets\newline (same experiment) & -- / 400 & 0.70, 1.00 & 1 & $1,\infty$ & 4 & 1 &
0.250, 1.000 ($q=0.70$)\newline 1.000, 1.414 ($q=1.00$) &
Opposite signs at the endpoints of each prespecified bracket. \\ \hline
Evidence versus score & -- / 1,600 & 0.70 & 1 & $\infty$ & 4 & 1 &
0.177, 0.595, 1.000 &
Replay reversal at 0.595 and agreement at the low and high anchors. \\ \hline
Exploitation prediction & 400 / 1,600 & 0.70 & 1 & $\infty$ & 4 & 1 &
0.177, 0.707 &
Concordance drop and replay reversal at 0.707 and agreement at 0.177. The product rule's gain at 0.707 is a secondary test. \\ \hline
\end{tabularx}
\end{table*}

The comparisons in Table~\ref{tab-3} use a relative-cost grid $\kappa=2^{k/4}$. The cost and score-quality contrasts weight both development limits equally. Earlier instances informed the evidence-versus-score design, and separate exploratory instances informed the prediction experiment's cost choices. We average paired instance-level contrasts. When a contrast combines configurations or cost levels, each bootstrap resample carries the same instance across the relevant cells. Intervals for the prediction experiment condition on the fitted scorer, since the bootstrap resamples only evaluation instances.

\textbf{Convention.} Two kinds of analysis appear in this paper. \emph{Registered} tests are those in Table~\ref{tab-3} and the policy-selection tests of Section~\ref{sec-6-1}, together with their prespecified secondary tests. Versioned specifications fixed their configurations, comparisons, decision rules, and seeds before the runs~\cite{ref-artifact}. Every other analysis is \emph{descriptive}. Run-in headings mark the registered and prespecified results, and unmarked analyses are descriptive. Unless stated otherwise, brackets and bands give pointwise 95\% percentile intervals from 5,000 paired bootstrap resamples, with no adjustment for multiple comparisons. The registered main loss comparisons require the interval to exclude zero in the specified direction and the absolute point estimate to reach a prespecified practical-magnitude bar of 2\% of the mean status-quo loss under its own response. If that reference loss is below 1\% of the mean loss with no patching and development at publication for every vulnerability, the main comparison reports the absolute difference and fails the relative-loss criterion. The concordance test instead requires a positive interval and a point estimate of at least $0.02$. The product rule's secondary test uses direction alone.

\subsubsection{Implementation checks}\label{sec-3-4-5}

We checked period-level loss, development costs, and listing dates against an independent implementation on small instances. Frozen runs also used hand-worked tests, verified that $\beta=0$ makes defender losses invariant to development cost at fixed inputs, and recovered the overlapping results from earlier frozen runs within their specified tolerances.

\section{Adoption beliefs and rule rankings}\label{sec-3-5}

This section relates the two evaluations analytically. Replay and policy-specific evaluation are the two ends of a family indexed by the adoption share the attacker believes, which gives bounds on the gap between them and conditions for a change of ranking.

Fix an instance and configuration with an endogenous catalog. Let $\ell_\pi(S)$ and $\ell_{\pi_0}(S)$ be the losses under full adoption of rule $\pi$ and the status quo, respectively, for the same plan $S$. Write $\Delta_\pi(S)=\ell_\pi(S)-\ell_{\pi_0}(S)$ and let $C(S)$ denote plan development cost.

\begin{lemma}[Scaling]\label{lem:adoption-scaling}
If a share $a$ adopts $\pi$, the population loss conditional on plan $S$ is $(1-a)\ell_{\pi_0}(S)+a\ell_\pi(S)$.
\end{lemma}

\begin{IEEEproof}
The plan fixes the listing calendar. Scaling a group's initial exposure and budget by the same factor scales baseline repair, proportional allocation, and the score-ordered queue by that factor. Inducting over periods gives the same scaling for remaining exposure and loss. The two groups contribute their respective shares of the corresponding full-adoption losses. See Appendix~\ref{app:adoption-scaling}.
\end{IEEEproof}

Now let actual adoption be complete and let $a\in[0,1]$ be the share the attacker believes has adopted $\pi$. With $\pi$ fixed, write $\sigma_a$ for the attacker's response distribution. For finite $\beta$,
\begin{equation}
\begin{aligned}
u_{\pi,a}(S) &= \ell_{\pi_0}(S)+a\Delta_\pi(S)-C(S),\\
\sigma_a(S) &=
\frac{\exp\!\bigl(\beta u_{\pi,a}(S)/v_{\mathrm{unit}}\bigr)}
     {\sum_{S'}\exp\!\bigl(\beta u_{\pi,a}(S')/v_{\mathrm{unit}}\bigr)},\\
V_\pi(a) &= \mathbb E_{\sigma_a}[\ell_\pi].
\end{aligned}
\label{eq:belief-response}
\end{equation}
The attacker evaluates payoffs at believed adoption, while $V_\pi(a)$ measures defender loss at actual full adoption.

\begin{proposition}[Endpoints]
Fixed-response replay gives $V_\pi(0)=J(\pi\mid\pi_0)$, and policy-specific evaluation gives $V_\pi(1)=J(\pi\mid\pi)$.
\end{proposition}

\begin{IEEEproof}
At $a=0$, both rules face the status-quo response. At $a=1$, each rule faces its own response. In both cases, $V_\pi$ averages the same full-adoption loss as in Section~\ref{sec-3-4-1}.
\end{IEEEproof}

\begin{proposition}[Selection of plans]
Under logit response,
\begin{equation}
\frac{d}{da}\mathbb E_{\sigma_a}[\Delta_\pi]
= \frac{\beta}{v_{\mathrm{unit}}}
  \operatorname{Var}_{\sigma_a}(\Delta_\pi) \ge 0.
\label{eq:belief-variance}
\end{equation}
Under exact best response, if $S_0,S_1$ are the chosen plans at the two endpoints, then $\Delta_\pi(S_1)\ge\Delta_\pi(S_0)$.
\end{proposition}

\begin{IEEEproof}
Differentiating~\eqref{eq:belief-response} gives
\[
\partial_a\log\sigma_a(S)=\frac{\beta}{v_{\mathrm{unit}}}
  \bigl(\Delta_\pi(S)-\mathbb E_{\sigma_a}[\Delta_\pi]\bigr).
\]
Multiplying by $\sigma_a(S)\Delta_\pi(S)$ and summing over plans yields~\eqref{eq:belief-variance}. For exact best response, add $u_{\pi,0}(S_0)\ge u_{\pi,0}(S_1)$ and $u_{\pi,1}(S_1)\ge u_{\pi,1}(S_0)$.
\end{IEEEproof}

\begin{proposition}[Two-sided endpoint bound]
For $0<\beta<\infty$, let $H(\sigma)=-\sum_S\sigma(S)\log\sigma(S)$, with $0\log 0=0$. Define
\[
\begin{aligned}
M_\pi={}&\mathbb E_{\sigma_1}[C]-\mathbb E_{\sigma_0}[C]\\
 &-\frac{v_{\mathrm{unit}}}{\beta}
       \bigl[H(\sigma_1)-H(\sigma_0)\bigr].
\end{aligned}
\]
Then
\begin{equation}
\begin{aligned}
M_\pi &\le V_\pi(1)-V_\pi(0)\\
      &\le M_\pi+\mathbb E_{\sigma_1}[\Delta_\pi]-\mathbb E_{\sigma_0}[\Delta_\pi],
\end{aligned}
\label{eq:belief-entropy-bound}
\end{equation}
and $\mathbb E_{\sigma_1}[\Delta_\pi]-\mathbb E_{\sigma_0}[\Delta_\pi]$ lies between zero and $\max_S\Delta_\pi(S)-\min_S\Delta_\pi(S)$. For exact best response,
\[
\begin{aligned}
C(S_1)-C(S_0)
  &\le \ell_\pi(S_1)-\ell_\pi(S_0)\\
  &\le C(S_1)-C(S_0)\\
  &\phantom{\le {}}+\Delta_\pi(S_1)-\Delta_\pi(S_0).
\end{aligned}
\]
\end{proposition}

\begin{IEEEproof}
At either endpoint the logit response maximizes expected net payoff plus $(v_{\mathrm{unit}}/\beta)H(\sigma)$. The objective equals the log normalizer times $v_{\mathrm{unit}}/\beta$ minus $(v_{\mathrm{unit}}/\beta)$ times the relative entropy of $\sigma$ with respect to the endpoint response. Comparing $\sigma_1$ with $\sigma_0$ in the endpoint-one objective $\mathbb E_\sigma[\ell_\pi-C]+(v_{\mathrm{unit}}/\beta)H(\sigma)$ yields the lower bound.

Since $\sigma_0$ maximizes $\mathbb E_\sigma[\ell_{\pi_0}-C]+(v_{\mathrm{unit}}/\beta)H(\sigma)$, comparing it with $\sigma_1$ gives $\mathbb E_{\sigma_1}[\ell_{\pi_0}]-\mathbb E_{\sigma_0}[\ell_{\pi_0}]\le M_\pi$. Add $\mathbb E_{\sigma_1}[\Delta_\pi]-\mathbb E_{\sigma_0}[\Delta_\pi]$ and use $\ell_\pi=\ell_{\pi_0}+\Delta_\pi$. The difference of expectations is nonnegative by~\eqref{eq:belief-variance} and cannot exceed the range of $\Delta_\pi$.

For exact best response, $S_1$ maximizes $\ell_\pi-C$ and $S_0$ maximizes $\ell_{\pi_0}-C$. Comparing the two plans in each objective gives the lower and upper bounds, respectively, using the same substitution for $\ell_\pi$.
\end{IEEEproof}

The gap between policy-specific and replay loss is pinned within a window whose width is at most the range of the rule's effect across plans. The attacker's net payoff is defender loss minus development cost, and the defender's payoff is minus its loss. When $C$ is zero for all plans, the game is zero-sum. Under exact best response, the lower bound is then zero, so replay gives a lower bound on each rule's policy-specific loss. More generally, any decrease under exact best response is bounded above by the savings in development cost, $C(S_0)-C(S_1)$.

\begin{proposition}[Defender loss and rule ranking]\label{prop:belief-rule-ranking}
For finite $\beta$,
\begin{equation}
\frac{d}{da}V_\pi(a)=\frac{\beta}{v_{\mathrm{unit}}}
\operatorname{Cov}_{\sigma_a}(\ell_\pi,\Delta_\pi).
\label{eq:belief-loss-slope}
\end{equation}
The difference $D(a)=V_{\pi_E}(a)-V_{\pi_S}(a)$ is continuous, as is its mean over instances. Opposite endpoint signs imply at least one change in ranking. At fixed finite $\beta>0$, expected losses at every $a$ tend to zero as the development-cost multiplier tends to infinity. With exact best response, they are zero at sufficiently high cost.
\end{proposition}

\begin{IEEEproof}
Apply the logit derivative identity with $\ell_\pi(S)$ in place of the multiplier $\Delta_\pi(S)$. The covariance can have either sign. The probabilities in~\eqref{eq:belief-response} are continuous, so the sign-change claim follows from the intermediate value theorem. At sufficiently high cost, net payoffs for plans with completed development tend to minus infinity, while plans without completed development have zero payoff and loss. Logit response assigns the former vanishing probability. Exact best response selects a zero-loss plan once all plans with completed development have negative payoffs.
\end{IEEEproof}

\begin{figure*}[!t]
\centering
\includegraphics[width=.84\textwidth]{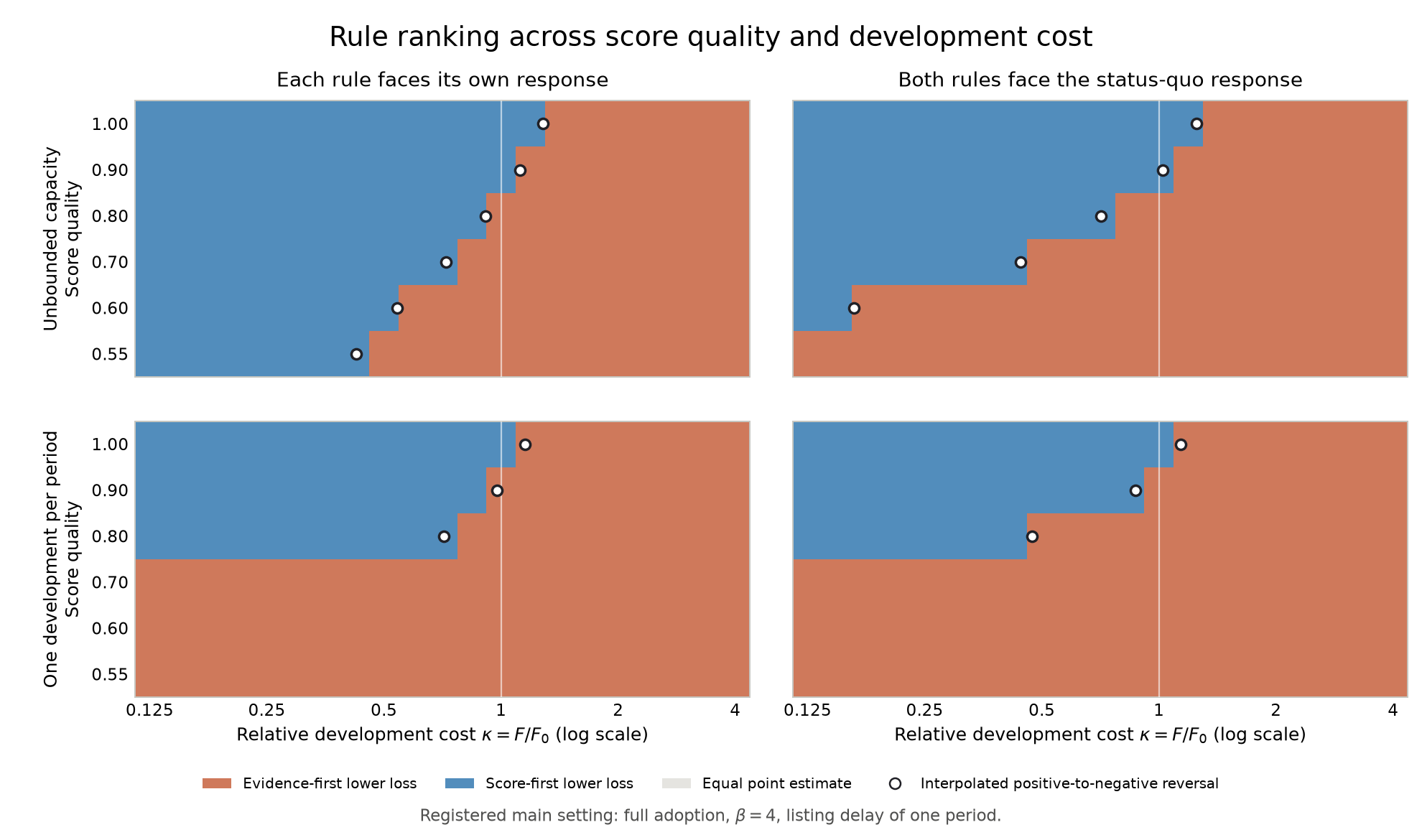}
\caption{Descriptive rankings of evidence-first and score-first across costs and score qualities in the registered main setting. The left column uses policy-specific evaluation and the right column uses fixed-response replay. The upper row has $K=\infty$ and the lower row has $K=1$. Colors indicate which rule has lower estimated loss, with a separate color for equal estimates. Circles indicate interpolated positive-to-negative crossings~\cite{ref-artifact}. The plotted cells and interpolations are descriptive.}\label{fig-1}
\end{figure*}

\refstepcounter{example}\label{ex:two-vulnerabilities}
\noindent\textbf{Example \theexample\ (two vulnerabilities, exact best response).}
Consider an exact publication score, exact best response, unlimited development per period, full adoption, listing delay $L=1$, and horizon $T=8$. Vulnerabilities $A$ and $B$ are both published in period 0 with development difficulty 1. Vulnerability $A$ has loss weight 2 and 40 exposed instances, and it has no remedy, so it is never listed. Vulnerability $B$ has loss weight 1 and 100 exposed instances, and it has a remedy. Thus $v_{\mathrm{unit}}=720$ and the per-period patch budget is 7. Table~\ref{tab:two-vulnerabilities} gives the plan losses and best responses. Exact thresholds are the gross loss differences between successive plans divided by $720F_0$, using unrounded losses. The supplementary artifact includes the unit test \nolinkurl{tests/test_signal_feedback_two_vulnerability_example.py}~\cite{ref-artifact}.

\begin{table}[!t]
\caption{Plan losses, exact best responses, and evidence-first minus score-first loss gaps for Example~\ref{ex:two-vulnerabilities}. Threshold labels are approximate. Rows between thresholds give open intervals. Ties follow the plan-index rule.}
\label{tab:two-vulnerabilities}
\centering
\ruledtablestyle
\begin{tabularx}{\columnwidth}{|L|R|R|R|R|}
\hline
& \multicolumn{3}{c|}{\textbf{Plan loss}} & \\
\cline{2-4}
\textbf{Plan} & $\boldsymbol{\pi_0}$ & $\boldsymbol{\pi_E}$ & $\boldsymbol{\pi_S}$ & \textbf{Cost$/F$} \\ \hline
$\emptyset$ & 0 & 0 & 0 & 0 \\ \hline
$\{A\}$ & 228.8 & 228.8 & 228.8 & 720 \\ \hline
$\{B\}$ & 594.6 & 574.9 & 650.5 & 720 \\ \hline
$\{A,B\}$ & 935.1 & 954.8 & 879.2 & 1440 \\ \hline
\end{tabularx}

\smallskip
\begin{tabularx}{\columnwidth}{|l|C|C|C|R|r|}
\hline
& \multicolumn{3}{c|}{\textbf{Best response to}} & \multicolumn{2}{c|}{\textbf{Loss gap}} \\
\cline{2-4}\cline{5-6}
$\boldsymbol{\kappa=F/F_0}$ & $\boldsymbol{\pi_0}$ & $\boldsymbol{\pi_E}$ & $\boldsymbol{\pi_S}$
    & \textbf{Replay} & \textbf{Policy-spec.} \\ \hline
$<0.267$      & $\{A,B\}$ & $\{A,B\}$ & $\{A,B\}$ & $+75.6$ & $+75.6$ \\ \hline
$0.267$--$0.398$ & $\{A,B\}$ & $\{A,B\}$ & $\{B\}$   & $+75.6$ & $+304.3$ \\ \hline
$0.398$--$0.444$ & $\{B\}$   & $\{A,B\}$ & $\{B\}$   & $-75.6$ & $+304.3$ \\ \hline
$0.444$--$0.671$ & $\{B\}$   & $\{B\}$   & $\{B\}$   & $-75.6$ & $-75.6$ \\ \hline
$0.671$--$0.695$ & $\{B\}$   & $\emptyset$ & $\{B\}$ & $-75.6$ & $-650.5$ \\ \hline
$0.695$--$0.759$ & $\emptyset$ & $\emptyset$ & $\{B\}$ & $0$ & $-650.5$ \\ \hline
$>0.759$      & $\emptyset$ & $\emptyset$ & $\emptyset$ & $0$ & $0$ \\ \hline
\end{tabularx}
\end{table}

The marginal gross return from adding $A$ to $\{B\}$ is 228.8 under score-first, 340.6 under the status quo, and 379.9 under evidence-first. Score-first repairs the higher-weight $A$ first, while evidence-first sends the budget to $B$ once it is listed and leaves the unlistable $A$ exposed to more loss. As development cost rises, the attacker therefore drops $A$ first under score-first, then under the status quo, and finally under evidence-first. Between the last two of these thresholds, replay evaluates both rules on the status-quo plan $\{B\}$, where evidence-first protects $B$ better, but evidence-first still faces $\{A,B\}$ under its own response, so the evaluations rank the rules in opposite orders.

This constructed instance uses exact best response and gives equal gaps at low and high costs and opposite rankings on an intermediate interval. With logit response and positive finite sensitivity, the response in this instance depends on the rule at every finite cost, so such low-cost agreement is not guaranteed in general. The registered intermediate-cost comparison in Section~\ref{sec-5-1} gives opposite endpoint signs for the sample-mean loss gap (Fig.~\ref{fig:adoption-belief}, left).

\begin{proposition}[Small adoption]\label{prop:small-adoption}
For finite $\beta$, let the actual adoption share $\alpha$ be known to the attacker. Conditional on plan $S$, population loss is $\ell_{\pi_0}+\alpha\Delta_\pi$ and adopters' own loss is $\alpha\ell_\pi$. The response is $\sigma_\alpha$ from~\eqref{eq:belief-response} at $a=\alpha$. For two rules $\pi_E,\pi_S$, write $\Delta_E,\Delta_S$ for their effects and $\sigma_0$ for the status-quo response. The replay gap in population loss is exactly $\alpha\mathbb E_{\sigma_0}[\Delta_E-\Delta_S]$. The policy-specific population loss gap $D_{\mathrm{adopt}}(\alpha)$ satisfies
\[
\begin{aligned}
\lim_{\alpha\to 0}\frac{D_{\mathrm{adopt}}(\alpha)}{\alpha}
={}&\mathbb E_{\sigma_0}[\Delta_E-\Delta_S]\\
 &+\frac{\beta}{v_{\mathrm{unit}}}
   \operatorname{Cov}_{\sigma_0}(\ell_{\pi_0},\Delta_E-\Delta_S).
\end{aligned}
\]
Adopters' own loss per unit of adopting exposure, $\mathbb E_{\sigma_\alpha}[\ell_\pi]$, tends to its replay value $\mathbb E_{\sigma_0}[\ell_\pi]$.
\end{proposition}

\begin{IEEEproof}
Under the common replay response the status-quo term cancels. For policy-specific loss, differentiate $\mathbb E_{\sigma_\alpha}[\ell_{\pi_0}+\alpha\Delta_\pi]$ at $\alpha=0$ by the logit identity. The direct term is $\mathbb E_{\sigma_0}[\Delta_\pi]$ and the change in response contributes $(\beta/v_{\mathrm{unit}})\operatorname{Cov}_{\sigma_0}(\ell_{\pi_0},\Delta_\pi)$. Subtract the two derivatives, using equality of population losses at $\alpha=0$. The adopters' convergence follows from continuity of $\sigma_\alpha$.
\end{IEEEproof}

For adopters' own loss per unit of exposure, replay and policy-specific evaluation converge as adoption shrinks. The rankings eventually agree when the limiting replay gap is nonzero. For population loss, the change in the attacker's response also affects non-adopters. This effect can change the sign of the leading-order population gap even when adoption is small. In the descriptive 30\%-adoption comparisons in Section~\ref{sec-5}, the point estimates still have opposite signs, with both relative magnitudes below 2\% for probability-only ranking.

\section{Which patching rule has lower loss?}\label{sec-4}

\subsection{Registered main results}\label{sec-4-1}

With an exact publication score, score-first incurred less loss over the prespecified low-cost aggregate, while evidence-first incurred less over the high-cost aggregate. Score quality could also reverse the comparison at a fixed cost. These results compare the proportional evidence-first rule from Section~\ref{sec-3-2} with score-first, each evaluated under its own attacker response.

The main setting uses full adoption, a one-period listing delay, and $\beta=4$. The registered contrasts weight the development limits $K=1$ and $K=\infty$ equally. In Table~\ref{tab-4}, $D_{\mathrm{adopt}}$ is evidence-first loss minus score-first loss, in loss points. Relative magnitudes express absolute point estimates as percentages of the corresponding status-quo losses. The paired 95\% bootstrap intervals are in loss points.

\begin{table*}[!t]
\caption{Registered evidence-first versus score-first contrasts across development costs and publication-score qualities. Positive values favor score-first.}\label{tab-4}
\centering
\ruledtablestyle
\begin{tabular}{|l|c|c|c|c|}
\hline
\textbf{Contrast} & \textbf{Publication-score quality $q$} & \textbf{Relative cost $\kappa$} & \textbf{Relative magnitude} & \textbf{$D_{\mathrm{adopt}}$ [95\% interval]} \\ \hline
Low-cost aggregate & 1.00 & 0.250, 0.354, 0.500 & 20.7\% & +159.1 [146.2, 172.6] \\ \hline
High-cost aggregate & 1.00 & 1.414, 2.000 & 17.7\% & \ensuremath{-}13.7 [\ensuremath{-}20.3, \ensuremath{-}7.2] \\ \hline
Fixed cost, exact score & 1.00 & 0.707 & 21.3\% & +110.4 [95.3, 126.1] \\ \hline
Fixed cost, noisy score & 0.55 & 0.707 & 13.2\% & \ensuremath{-}91.4 [\ensuremath{-}107.0, \ensuremath{-}75.4] \\ \hline
\end{tabular}
\end{table*}

All four contrasts met the registered criteria. The absolute gap was smaller in the high-cost aggregate than in the low-cost aggregate. Development was rarer and the status-quo loss was lower in the high-cost comparison, so the gap was still 17.7\% of that comparison's status-quo loss (13.7 points).

\subsection{Prespecified comparison of reversal intervals}\label{sec-4-2}

The additional registered test compared the signs of $D_{\mathrm{adopt}}$ at four fixed endpoints. All four met the direction and magnitude criteria in Table~\ref{tab-5}.

\begin{table*}[!t]
\caption{Additional registered contrasts at the endpoints of the rule-reversal intervals. Positive values favor score-first.}\label{tab-5}
\centering
\ruledtablestyle
\begin{tabular}{|l|c|c|c|c|}
\hline
\textbf{Endpoint} & \textbf{$q$} & \textbf{$\kappa$} & \textbf{Relative magnitude} & \textbf{$D_{\mathrm{adopt}}$ [95\% interval]} \\ \hline
Lower endpoint & 0.70 & 0.250 & 4.5\% & +41.1 [28.0, 54.6] \\ \hline
Upper endpoint & 0.70 & 1.000 & 19.8\% & \ensuremath{-}85.3 [\ensuremath{-}103.1, \ensuremath{-}67.3] \\ \hline
Lower endpoint & 1.00 & 1.000 & 10.6\% & +31.7 [18.1, 46.3] \\ \hline
Upper endpoint & 1.00 & 1.414 & 11.2\% & \ensuremath{-}13.7 [\ensuremath{-}22.7, \ensuremath{-}4.1] \\ \hline
\end{tabular}
\end{table*}

Under equal weighting of the two development limits, the endpoints bracket at least one change in rule order between $\kappa=0.250$ and $\kappa=1$ at $q=0.70$, and between $\kappa=1$ and $\kappa=1.414$ at $q=1.00$. The test establishes the order of those brackets. The estimate at the lower endpoint for $q=0.70$ was +85.8 [72.2, 100.0] with unlimited development capacity and \ensuremath{-}3.6 [\ensuremath{-}16.9, 10.0] with $K=1$.

\subsection{Descriptive comparisons}\label{sec-4-3}

Figure~\ref{fig-1} shows the signs of the point estimates across the grid by score quality, cost, and development capacity.

\textbf{Crossings on the full grid.} Under unlimited development capacity, the interpolated crossing for policy-specific loss moved from $\kappa=0.42$ [0.32, 0.51] at $q=0.55$ to $\kappa=1.28$ [1.17, 1.39] at $q=1.00$. The intervals condition on bootstrap resamples containing a crossing.

\textbf{Evidence-plus-score.} The rule that ranks listed vulnerabilities by publication score before spending any remaining budget on unlisted ones outperformed proportional evidence-first in 201 of 252 cells by pointwise intervals, and underperformed it in two. In the comparison with score-first, 154 intervals favored evidence-plus-score and 80 favored score-first. At low costs and high score quality, score-first could outperform even the rule that uses both signals. The registered reversal ranks two rules. The comparison with evidence-plus-score shows why a choice among all three requires the additional loss estimates.

\textbf{Sensitivity to model inputs.} Reapplying the main criteria in the descriptive variants showed that the high-cost aggregate missed the criterion under partial adoption, $\beta=1$, and the larger patch budget. The low-cost aggregate and the fixed-cost contrast with the noisy score still met the criteria in each of the eight variants examined~\cite{ref-artifact}. The registered comparisons concern the specified synthetic costs and scores.

\section{When replay reverses the rule ranking}\label{sec-5}

The rule comparisons in Section~\ref{sec-4} let each policy induce its own attacker response. When both policies instead face the status-quo plan distribution, the size of the loss gap can change. In the prespecified secondary analysis of the cost and score-quality experiment, the interval for $D_{\mathrm{adopt}}-D_{\mathrm{replay}}$ excluded zero in seven of the eight aggregate contrasts. For example, the low-cost aggregate gave +70.3 [66.1, 74.7] loss points on this contrast, moving from a replay gap of +88.8 [79.1, 99.0] to a policy-specific gap of +159.1 [146.2, 172.6]. None of the eight aggregate contrasts established opposite rule rankings. The full grid supplied candidate conditions for a separate test on fresh instances.

\begin{figure*}[!t]
\centering
\includegraphics[width=.92\textwidth]{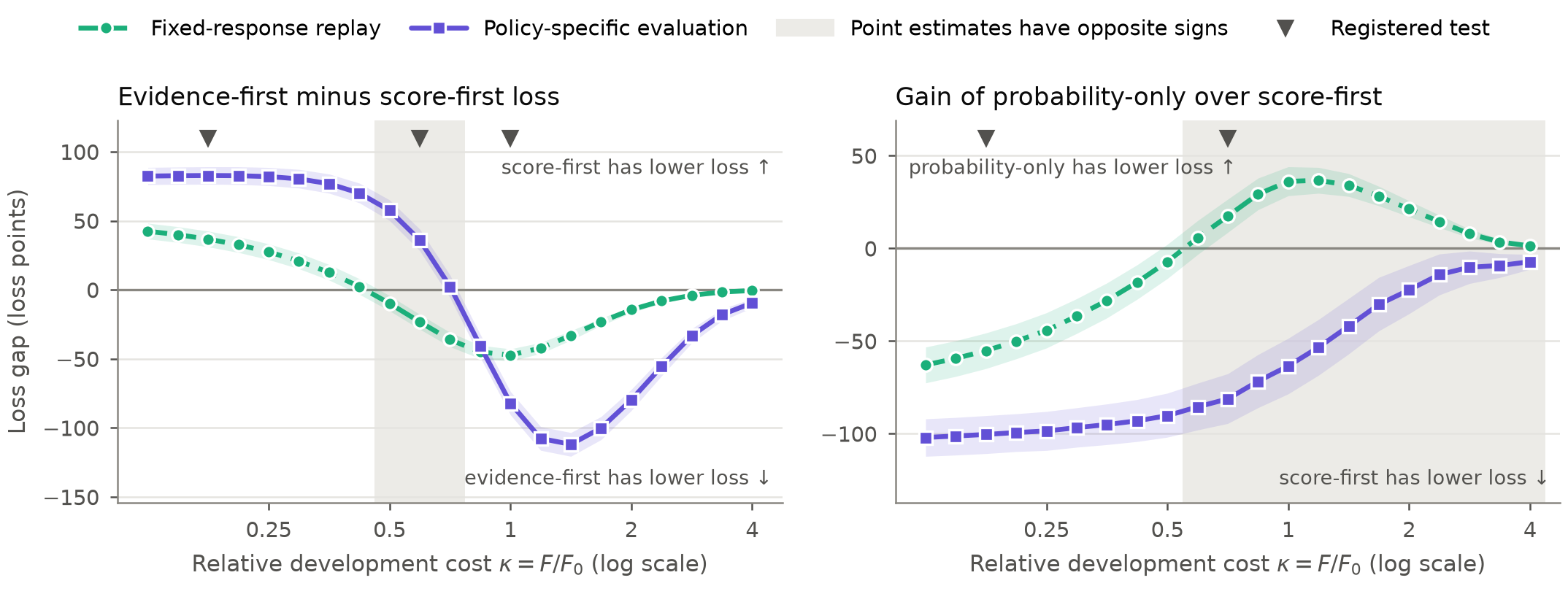}
\caption{Loss gaps under fixed-response replay and under policy-specific evaluation at each of the 21 relative costs, in the registered configurations. Left: evidence-first minus score-first loss. Right: gain of probability-only ranking over score-first. Bands are 95\% intervals. Shading marks the cost cells in which the two point estimates have opposite signs, and triangles mark the costs of the registered tests~\cite{ref-artifact}.}\label{fig-2}
\end{figure*}

Figure~\ref{fig-2} shows both evaluations over the 21 relative costs of the two experiments reported in this section. The registered tests concern the marked costs. The remaining cells are descriptive.

\begin{figure*}[!t]
\centering
\includegraphics[width=.92\textwidth]{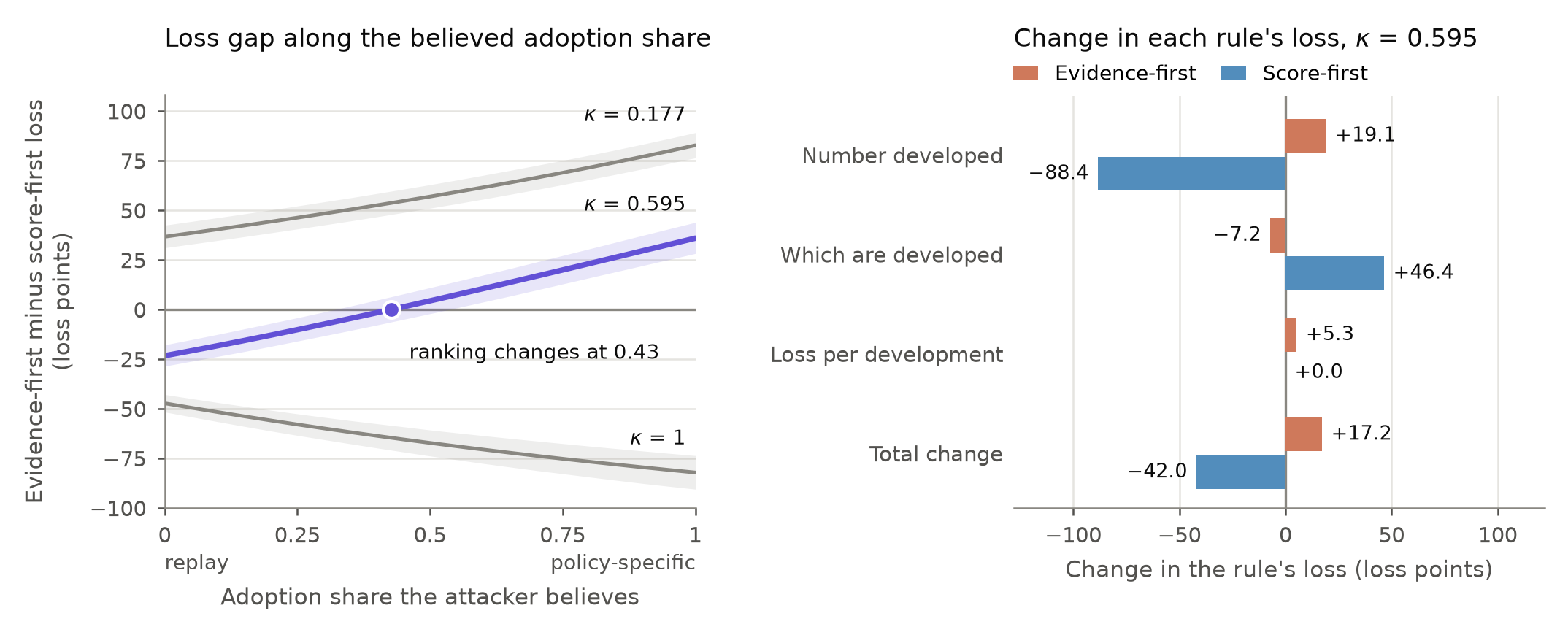}
\caption{Believed adoption and the changes in loss between evaluation methods. The left panel shows $D(a)$ at the three registered costs, with 95\% bands and the estimated sign change marked at $a=0.43$. The right panel shows the count, composition, conditional-loss, and total changes from replay to policy-specific evaluation for each rule at $\kappa=0.595$~\cite{ref-artifact}.}\label{fig:adoption-belief}
\end{figure*}

\subsection{Evidence-first versus score-first}\label{sec-5-1}

\textbf{Registered replay reversal.} At the prespecified intermediate cost, fixed-response replay favored evidence-first by 3.0\% of status-quo loss (23.3 points), and policy-specific evaluation favored score-first by 4.7\% (36.0 points). With full adoption, these are the $a=0$ and $a=1$ belief endpoints, respectively (Section~\ref{sec-3-5}). Both evaluations favored score-first at the low-cost anchor and evidence-first at the high-cost anchor. All three tests met their registered criteria in the configuration with $q=0.70$, full adoption, unlimited development capacity per period, $\beta=4$, and a one-period listing delay (Table~\ref{tab-6}).

\begin{table}[!t]
\caption{Registered replay and policy-specific contrasts for evidence-first versus score-first. Positive values favor score-first.}\label{tab-6}
\centering
\ruledtablestyle
\begin{tabularx}{\columnwidth}{|P{.19\columnwidth}|c|C|C|}
\hline
\textbf{Cost condition} & \textbf{$\kappa$} & \textbf{$D_{\mathrm{replay}}$\newline [95\% interval]} & \textbf{$D_{\mathrm{adopt}}$\newline [95\% interval]} \\ \hline
Low anchor & 0.177 & +36.7 [31.1, 42.5] & +82.8 [76.5, 89.1] \\ \hline
Intermediate & 0.595 & \ensuremath{-}23.3 [\ensuremath{-}28.5, \ensuremath{-}17.8] & +36.0 [28.2, 43.9] \\ \hline
High anchor & 1.000 & \ensuremath{-}47.2 [\ensuremath{-}51.6, \ensuremath{-}42.8] & \ensuremath{-}82.1 [\ensuremath{-}90.5, \ensuremath{-}73.6] \\ \hline
\end{tabularx}
\end{table}

The signs bracket a change in rule order between $\kappa=0.177$ and $\kappa=0.595$ in replay, and between $\kappa=0.595$ and $\kappa=1$ under policy-specific responses~\cite{ref-artifact}. The registered result establishes a case in which replay favors the rule with higher loss under policy-specific evaluation.

\textbf{Prespecified secondary test of development counts.} At the intermediate cost, the attacker developed an expected 3.68 vulnerabilities per instance under evidence-first and 3.19 under score-first. The paired difference was +0.485 [0.466, 0.505]. Development counts changed with the rule.

\textbf{Believed-adoption curve.} We used the 1,600 instances of the registered experiment to compute the mean evidence-first minus score-first loss gap $D(a)$ at $a=0,0.05,\ldots,1$ for the three registered costs, with actual adoption fixed at one (Section~\ref{sec-3-5}). The curve and loss split use the same paired bootstrap. At all three costs, both endpoint estimates and intervals reproduce the registered values in Table~\ref{tab-6} exactly. At $\kappa=0.595$, the gap rises monotonically over the 21 grid points from $-23.3$ [$-28.5$, $-17.8$] at $a=0$ to $+36.0$ $[28.2,\,43.9]$ at $a=1$. It is $-1.6$ [$-7.8$, 4.8] at $a=0.4$ and $+4.4$ [$-2.1$, 11.0] at $a=0.5$. The mean curve changes sign once at $a=0.43$ by linear interpolation, with a 95\% bootstrap interval of [0.32, 0.54] for the location. Each of the 5,000 resampled mean curves changes sign exactly once. In this configuration, the ranking flips once the attacker believes that roughly 43\% of organizations follow the new rule. At $\kappa=0.177$, the gap rises monotonically from $+36.7$ [31.1, 42.5] to $+82.8$ [76.5, 89.1]. At $\kappa=1$, it falls monotonically from $-47.2$ [$-51.6$, $-42.8$] to $-82.1$ [$-90.5$, $-73.6$]. The gap keeps its sign at both anchors. The curve can be non-monotone in general (Proposition~\ref{prop:belief-rule-ranking}) and was monotone on the grid at these three costs (Fig.~\ref{fig:adoption-belief}, left). The belief $a$ is an exogenous input, with no modeled reconnaissance or belief updating.

\textbf{Split of the loss change.} For a fixed rule $\pi$ and instance, let $p_i$ be the probability of developing vulnerability $i$ under a response and $\mu_i$ its mean loss given development. The expected count is $\nu=\sum_i p_i$, and $\omega_i=p_i/\nu$ is vulnerability $i$'s share of that count. Here $K=\infty$ makes development at publication feasible for every singleton plan, and finite $\beta$ assigns positive probability to each plan. Thus $p_i>0$ and $\nu>0$ under both responses. Superscripts 0 and 1 denote the status-quo and rule-specific responses, respectively. The loss change splits exactly into three margins.
\[
\begin{aligned}
&V_\pi(1)-V_\pi(0) \\
&\quad= \underbrace{(\nu^1-\nu^0)\sum_i \omega_i^0 \mu_i^0}_{\text{count}} \\
&\quad+ \underbrace{\nu^1\sum_i (\omega_i^1-\omega_i^0)\mu_i^0}_{\text{composition}} \\
&\quad+ \underbrace{\sum_i p_i^1(\mu_i^1-\mu_i^0)}_{\text{conditional loss}}.
\end{aligned}
\]
The count and composition margins value developments at the replay conditional means $\mu_i^0$, so the split depends on that reference. Table~\ref{tab:adoption-loss-split} reports the mean margins across instances. When the attacker accounts for score-first, it develops fewer vulnerabilities, reducing loss by 88.4 points. It keeps the more damaging ones, adding back 46.4 points, so score-first's loss under its own response is 42.0 points below replay. Since $K=\infty$ fixes completion at publication and score-first ignores the catalog, the loss from a developed vulnerability is independent of which others are developed. Its conditional-loss margin is exactly zero. Under evidence-first, the attacker develops more vulnerabilities, adding 19.1 loss points. Because the budget follows listings, the change in loss conditional on development adds another 5.3 points. The number developed drives the 59.2-point swing between replay and policy-specific evaluation at this cost. Composition offsets about half of the count margin, and loss per development contributes a small share (Fig.~\ref{fig:adoption-belief}, right).

\begin{table}[!t]
\caption{Split of the change from replay to policy-specific loss at $\kappa=0.595$, in loss points. The difference is evidence-first minus score-first. Brackets give 95\% intervals.}\label{tab:adoption-loss-split}
\centering
\ruledtablestyle
\setlength{\tabcolsep}{2pt}
\begin{tabular}{|>{\raggedright\arraybackslash}m{.135\columnwidth}|c|c|c|c|}
\hline
& \textbf{Count} & \textbf{Composition} & \stackedtablecell{\textbf{Conditional}\\\textbf{loss}} & \textbf{Total} \\ \hline
Score-first & \intervalentry{-88.4}{-92.5,\,-84.5} & \intervalentry{+46.4}{44.1,\,48.8} & $0.0$ & \intervalentry{-42.0}{-44.8,\,-39.4} \\ \hline
Evidence-first & \intervalentry{+19.1}{17.8,\,20.6} & \intervalentry{-7.2}{-7.8,\,-6.6} & \intervalentry{+5.3}{4.9,\,5.7} & \intervalentry{+17.2}{15.8,\,18.6} \\ \hline
Difference & \intervalentry{+107.6}{102.8,\,112.6} & \intervalentry{-53.6}{-56.3,\,-51.0} & \intervalentry{+5.3}{4.9,\,5.7} & \intervalentry{+59.2}{55.6,\,62.9} \\ \hline
\end{tabular}
\end{table}

At the low anchor ($\kappa=0.177$), the evidence-first minus score-first margins are $+84.7$ [82.1, 87.4] for count, $-41.7$ [$-43.5$, $-39.9$] for composition, and $+3.0$ [2.9, 3.2] for conditional loss, for a total of $+46.1$ [44.7, 47.4]. At the high anchor ($\kappa=1$), the margins are $-0.2$ [$-5.7$, 5.5] for count, $-36.5$ [$-39.3$, $-33.8$] for composition, and $+1.8$ [1.5, 2.2] for conditional loss, for a total of $-34.9$ [$-40.5$, $-29.3$]. The high-anchor count interval includes zero, and composition accounts for nearly all of the change.

\textbf{Full cost grid.} The two estimates had opposite signs in three of the 21 cost cells. Both pointwise intervals excluded zero in two of those three. At the intermediate cost, replay still favored evidence-first when both rules used the plan distribution induced by evidence-first (\ensuremath{-}12.3 [\ensuremath{-}17.7, \ensuremath{-}6.7]) or by score-first (\ensuremath{-}71.2 [\ensuremath{-}76.2, \ensuremath{-}66.0]). These contrasts use different replay references and give different effect magnitudes. In the capacity variants, the intermediate reversal persisted with limits of two or three developments per period. With a limit of one, the policy-specific gap was \ensuremath{-}35.4 [\ensuremath{-}42.9, \ensuremath{-}28.0] and the replay gap was \ensuremath{-}47.7 [\ensuremath{-}53.0, \ensuremath{-}42.3], both favoring evidence-first at that cost. Under partial adoption or a three-period listing delay, the signs still opposed at that cost, with at least one magnitude in each pair below 2\%.

\textbf{Check with evidence-plus-score.} Evidence-plus-score ($\pi_{ES}$) had lower policy-specific loss than proportional evidence-first in 201 of 252 cells by pointwise intervals (Section~\ref{sec-4}). We compared $\pi_{ES}$ with score-first on 800 fresh instances under the registered configuration. The intervals and cell counts are descriptive. Table~\ref{tab:evidence-plus-score-replay} shows opposite point-estimate signs in 5 of 21 cost cells, with both intervals excluding zero at 3 costs ($\kappa=0.354$, 0.420, and 0.500). From $\kappa=0.707$ upward both estimates are negative. On the same instances at $\kappa=0.595$, proportional evidence-first minus score-first loss was $-21.7$ [$-29.2$, $-13.9$] in replay and $+41.8$ [30.7, 53.3] in policy-specific evaluation, close to the registered $-23.3$ and $+36.0$. Under their own responses, evidence-plus-score had about 41 points less loss than proportional evidence-first at low cost, with the gap shrinking to zero by $\kappa=1.68$.

\begin{table}[!t]
\caption{Evidence-plus-score minus score-first loss, in points. Negative values favor evidence-plus-score. Brackets give 95\% intervals.}\label{tab:evidence-plus-score-replay}
\centering
\ruledtablestyle
\begin{tabularx}{\columnwidth}{|c|R|R|}
\hline
\textbf{$\kappa$} & \textbf{Replay} & \textbf{Policy-specific} \\ \hline
0.297 & $-5.3$ [$-12.2$, 1.7] & $+44.5$ [36.8, 52.6] \\ \hline
0.354 & $-11.9$ [$-18.9$, $-4.8$] & $+41.3$ [33.4, 49.6] \\ \hline
0.420 & $-20.1$ [$-27.0$, $-13.0$] & $+35.0$ [26.7, 43.6] \\ \hline
0.500 & $-29.6$ [$-36.5$, $-22.5$] & $+23.3$ [14.5, 32.4] \\ \hline
0.595 & $-39.7$ [$-46.6$, $-32.7$] & $+3.8$ [$-5.7$, 13.4] \\ \hline
\end{tabularx}
\end{table}

\subsection{Ranking by predicted exploitation probability}\label{sec-5-2}

The exploitation prediction experiment froze a predictor fitted on the model's status-quo exploitation probabilities and tested it on fresh instances. Both probability-only and score-first commit their full budget to a single patch queue. They differ in the signal used to rank that queue. This experiment shares the configuration from the registered evidence-first versus score-first reversal and tests a different intermediate cost, $\kappa=0.707$. The gain contrasts $G_{\mathrm{replay}}$ and $G_{\mathrm{adopt}}$ are defined in Section~\ref{sec-3-4-2}. With actual full adoption, they likewise give the gain at believed adoption shares $a=0$ and $a=1$, respectively. The intervals condition on the fitted predictor.

\textbf{Registered concordance result.} At $\kappa=0.707$, concordance of the same predictor with the induced exploitation probabilities was 0.653 under score-first and 0.568 under probability-only ranking. The paired drop of 0.0856 [0.0809, 0.0902] met the criterion for lower concordance. This is a change in the agreement of the frozen predictions with the probabilities they are evaluated against.

\textbf{Registered loss results.} At the intermediate cost, replay credited probability-only ranking with a gain of 2.6\% of status-quo loss (17.4 points) over score-first, the smallest relative magnitude among the four registered loss estimates. When each rule induced its own response, probability-only incurred additional loss equal to 11.9\% of status-quo loss (81.2 points). At the low-cost anchor, replay and policy-specific evaluation both favored score-first, with a larger loss gap in the latter comparison. Both tests met the registered criteria (Table~\ref{tab-7}).

\begin{table}[!t]
\caption{Registered gains of probability-only ranking relative to score-first. Positive values favor probability-only ranking.}\label{tab-7}
\centering
\ruledtablestyle
\begin{tabularx}{\columnwidth}{|P{.19\columnwidth}|c|C|C|}
\hline
\textbf{Cost condition} & \textbf{$\kappa$} & \textbf{$G_{\mathrm{replay}}$\newline [95\% interval]} & \textbf{$G_{\mathrm{adopt}}$\newline [95\% interval]} \\ \hline
Intermediate & 0.707 & +17.4 [8.6, 26.5] & \ensuremath{-}81.2 [\ensuremath{-}94.5, \ensuremath{-}67.6] \\ \hline
Low anchor & 0.177 & \ensuremath{-}55.2 [\ensuremath{-}64.8, \ensuremath{-}45.6] & \ensuremath{-}100.4\newline [\ensuremath{-}110.7, \ensuremath{-}90.2] \\ \hline
\end{tabularx}
\end{table}

\textbf{Prespecified secondary analyses.} At the intermediate cost, the rule that multiplies predicted probability by the publication score had a policy-specific gain over score-first of 1.1\% of status-quo loss (+7.8 [3.6, 12.0] points). This met the prespecified direction criterion for the product rule, with a magnitude below the 2\% bar for the main loss tests. The product rule's replay gain was also positive, +14.6 [11.7, 17.5]. In a separate control for probability-only ranking, replay on the score-first response still favored probability-only at the intermediate cost, by +38.4 [29.9, 47.3]. In that comparison, score-first faces its own response in both evaluations, so the reversal arises when the response to probability-only is allowed to change.

\begin{figure*}[!t]
\centering
\includegraphics[width=.92\textwidth]{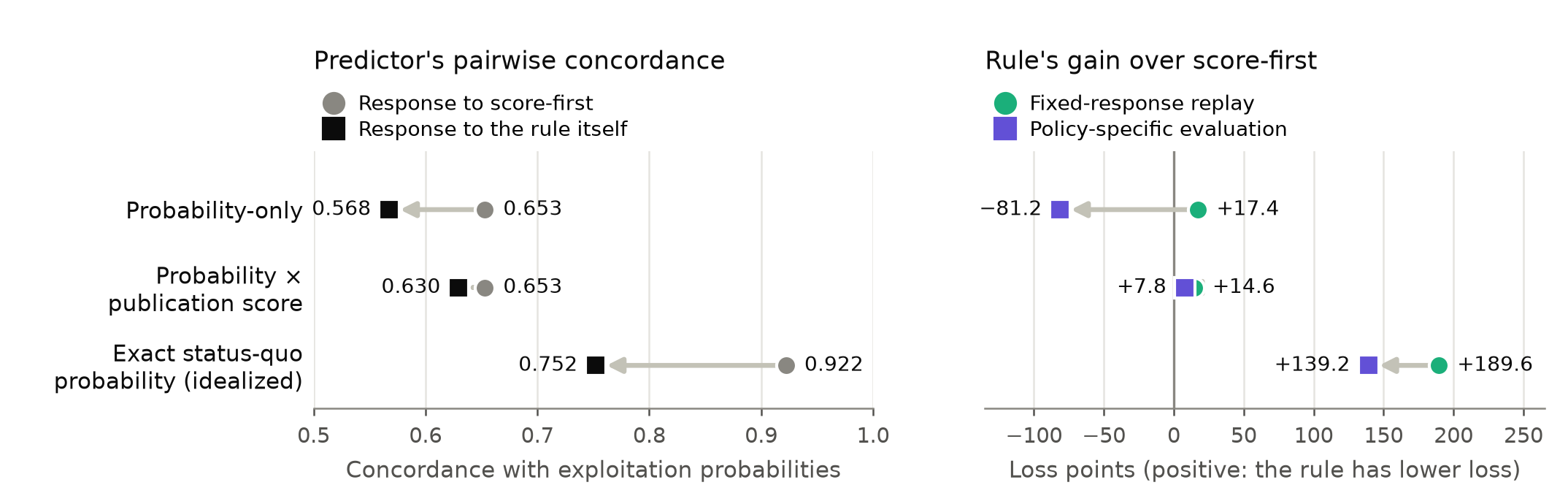}
\caption{The three rules that rank by a predicted exploitation probability, at the registered intermediate cost $\kappa=0.707$. Left: pairwise concordance of the predictor with exploitation probabilities under the response to score-first and under the response to the rule that uses it. Right: the rule's gain over score-first under fixed-response replay and under policy-specific evaluation, in loss points. Arrows run from the first to the second evaluation. Values are point estimates, with intervals in the text~\cite{ref-artifact}.}\label{fig-3}
\end{figure*}

\textbf{Full cost grid.} On the 21-cost grid, the point estimates for probability-only ranking had opposite signs in the two evaluations at 12 costs. Both intervals excluded zero in 11 of those cells, including two where the status-quo loss was too small for the relative criterion. At 30\% adoption, the intermediate reversal was +5.2 [2.6, 8.0] in replay and \ensuremath{-}13.1 [\ensuremath{-}17.2, \ensuremath{-}9.0] in policy-specific gain, with both magnitudes below 2\%. An idealized rule using exact status-quo exploitation probabilities provided a separate illustration. At the intermediate cost, its concordance dropped by 0.170 [0.164, 0.177] relative to evaluation under score-first's response, while its policy-specific gain over score-first was +139.2 [125.8, 152.7]~\cite{ref-artifact}. Lower concordance can therefore coexist with lower patching loss in this model. Figure~\ref{fig-3} places the three rules side by side: the predictor's concordance falls under each rule that uses it, while the gain over score-first is negative for probability-only ranking and positive for the other two.

\begin{figure*}[!t]
\centering
\includegraphics[width=.92\textwidth]{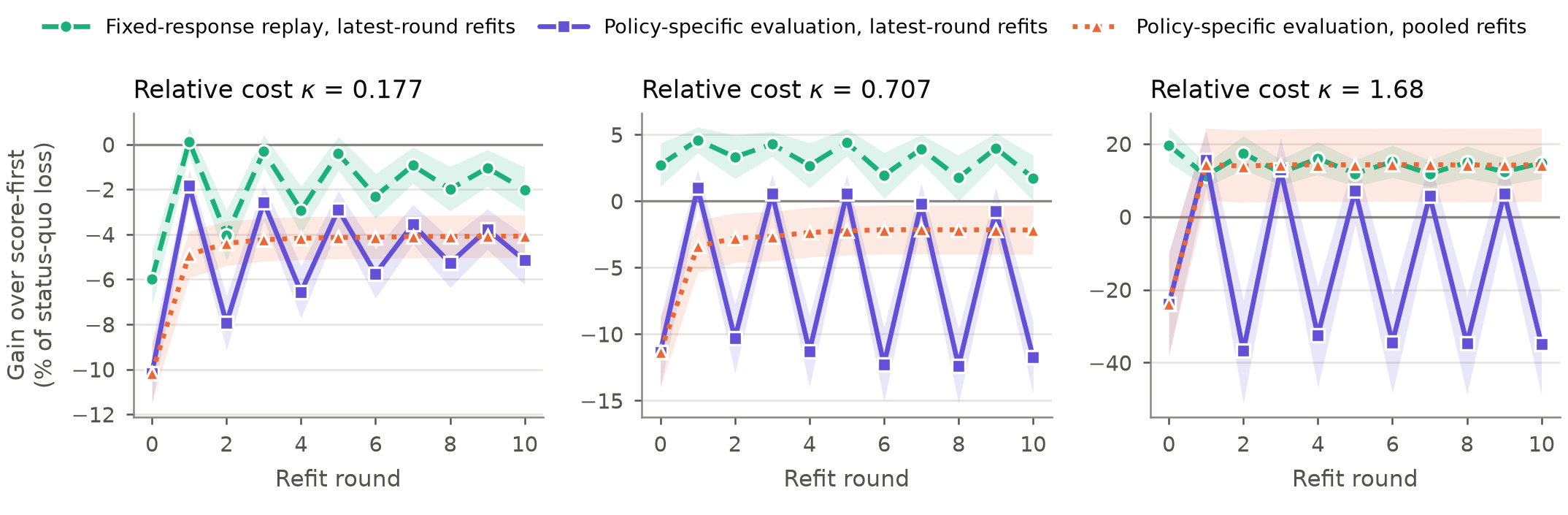}
\caption{Gain over score-first as a percentage of status-quo loss by refit round at $\kappa=0.177$, 0.707, and 1.68. Dashed green circles show fixed-response replay with latest-round refits. Solid violet squares show policy-specific evaluation with latest-round refits, and dotted orange triangles show policy-specific evaluation with pooled refits. Shaded bands are 95\% intervals conditional on each round's fitted predictor. Damped refits are not drawn: in every round their policy-specific gain is within 3.9 loss points, and within 1.0\% of status-quo loss, of the pooled refits (round 10 in Table~\ref{tab:refit-comparison})~\cite{ref-artifact}.}\label{fig:retraining-rounds}
\end{figure*}

\textbf{Variants.} All three registered criteria still held with 16 vulnerabilities, development limits of 2 or 3, a three-period listing delay, $\beta=16$ or exact best response, or $q=1.00$. They also held in each of the three instance-generation variants, which set the internet-facing share to 0.5, provided remedies for all vulnerabilities, or set the share without remedies to 0.4. Effect sizes differed across variants. With a development limit of 1, the low-cost criterion failed. With $\beta=1$, the reversal criterion failed. At 30\% adoption, only the concordance criterion held~\cite{ref-artifact}.

\textbf{Repeated refit.} We used the prediction experiment's configuration with 400 new fitting and 800 new evaluation instances. Round 0 fits on status-quo labels. Each of the ten subsequent latest-round refits uses the exploitation probabilities induced by deploying the previous round's rule. Gains are score-first loss minus the refitted rule's loss. Intervals in Fig.~\ref{fig:retraining-rounds} condition on each fitted predictor~\cite{ref-artifact}.

At $\kappa=0.707$ (status-quo loss 680.1), the policy-specific gain alternated. In even rounds it lay between $-84.3$ and $-70.1$, or 10.3\% to 12.4\% of status-quo loss. In odd rounds it lay between $-5.0$ and $+6.8$, with every interval including zero (Fig.~\ref{fig:retraining-rounds}). Replay gain stayed positive, between $+11.9$ and $+31.2$. Concordance under the predictor's own adoption stayed between 0.554 and 0.573, against 0.61 to 0.62 on the probabilities it was fitted to, and the largest parameter change between successive predictors (0.19 to 0.43) showed no sustained decline. The same alternation appeared at $\kappa=1.68$ (status-quo loss 131.4), with even-round gains between $-48.3$ and $-31.5$ and odd-round gains falling from $+20.4$ [10.4, 30.9] to between $+7.5$ and $+9.5$ in the last three rounds, whose intervals include zero. At $\kappa=0.177$ the gain was negative in every round and the swings shrank, from $-111.1$ [$-125.8$, $-96.5$] in round 0 to $-56.4$ [$-68.2$, $-44.5$] in round 10.

The coefficients show what alternates. From round 2 on at $\kappa=0.707$, even-round predictors put a standardized weight of 0.38--0.39 on log exposure and 0.07--0.10 on the log publication score. Odd-round predictors put 0.23--0.26 on exposure and 0.27--0.29 on the score, the relative-public-value coefficient rises from about 0.03 to 0.12--0.13, and the remedy coefficient changes sign between rounds ($+0.157$ in round 9, $-0.163$ in round 10). Each refit is trained on where development went under the previous rule. A predictor that ranks mostly by exposure leaves high-score vulnerabilities exposed longer, the development probabilities it induces are then better explained by the score, and the next fit leans on the score. That fit ranks more like score-first, the induced probabilities again line up with exposure, and the following fit swings back. We read the alternation as this two-step cycle between the defenders' ranking and the attacker's response. The coefficients are consistent with that reading and do not isolate it. At $\kappa=1.68$ the exposure and relative-public-value coefficients alternate in the same way. Replay favors probability-only in every round at both costs.

\textbf{Post hoc refit comparisons.} Pooled refits average the status-quo labels and all induced labels so far. Damped refits average the previous fitting targets and the newly induced labels with equal weights. Both reduce the large swings in gains (Fig.~\ref{fig:retraining-rounds}). At round 10 the maximum parameter change is below 0.003 for both updates at all three costs. The round-10 probability-only rules lose roughly 4\% and 2\% of status-quo loss relative to score-first at the low and intermediate costs and gain about 14\% at the sparse cost (Table~\ref{tab:refit-comparison}). At $\kappa=0.707$, pooling still gives a replay gain of $+36.2$ [27.8, 44.5] and a policy-specific gain of $-14.8$ [$-27.3$, $-2.2$]. With latest-round refits, the product rule gives positive round-10 gains at both higher costs and shows much smaller late-round swings. These finite-round results complement theoretical conditions for convergence in performative prediction~\cite{ref-4}. Smoothing the refits thus removes most of the alternation without removing the replay discrepancy at the intermediate cost. Among the four refits compared, only the product rule, which keeps the publication score in the ranking, has a round-10 gain with a nonnegative point estimate at all three costs.

\begin{table}[!t]
\caption{Round-10 policy-specific gains over score-first. Positive gains favor the refitted rule. Brackets give 95\% intervals. The product rule uses latest-round refits.}\label{tab:refit-comparison}
\centering
\ruledtablestyle
\begin{tabularx}{\columnwidth}{|l|C|C|C|}
\hline
\textbf{Refit} & $\boldsymbol{\kappa=0.177}$ & $\boldsymbol{\kappa=0.707}$ & $\boldsymbol{\kappa=1.68}$ \\ \hline
Latest round & \intervalentry{-56.4}{-68.2,\,-44.5} & \intervalentry{-79.8}{-98.6,\,-61.5} & \intervalentry{-45.9}{-64.3,\,-28.7} \\ \hline
Pooled & \intervalentry{-44.6}{-54.9,\,-34.3} & \intervalentry{-14.8}{-27.3,\,-2.2} & \intervalentry{+18.7}{5.5,\,31.9} \\ \hline
Damped & \intervalentry{-44.2}{-54.5,\,-33.8} & \intervalentry{-13.7}{-26.1,\,-1.2} & \intervalentry{+18.9}{5.7,\,32.1} \\ \hline
Product rule & \intervalentry{+0.3}{-1.6,\,2.4} & \intervalentry{+8.6}{3.6,\,13.8} & \intervalentry{+19.0}{11.8,\,26.4} \\ \hline
\end{tabularx}
\end{table}

\textbf{Synthesis.} Evidence-first versus score-first had opposite point-estimate signs in 3 of 21 cost cells, with both intervals excluding zero in 2, near costs at which the two rules have similar losses, with reversals also seen for evidence-plus-score. Probability-only versus score-first had opposite signs in 12 of 21 cells, with both intervals excluding zero in 11, a wider band. Probability-only ranks by development likelihood alone. The product rule also weights the predicted likelihood by the publication score, which accounts for the loss weight and can put different vulnerabilities at the head of the queue. This helps explain why the product rule's gain kept its sign across both evaluations at the tested intermediate~cost.

\section{Policy selection, attribute shares, and scalability}\label{sec-6}

\subsection{Policy selection}\label{sec-6-1}

Section~\ref{sec-5} compares the rankings of two fixed rules under replay and policy-specific responses. Here the decision is which budget split to select from a family of 11 two-layer policies, $\theta\in\{0,0.1,\ldots,1\}$. One selector minimizes mean replay loss against the status-quo plan distribution. The other minimizes mean loss under each split's own induced response. Both use the same information, patch budget, and selection instances. On a separate evaluation set, each selected policy faces its own induced response. The gain is the loss of the replay-selected policy minus the loss of the policy selected with policy-specific responses. It measures the value of accounting for response when choosing a budget split. Cost windows and policies are chosen on the selection set and held fixed during evaluation.

In this finite policy family, the leader's commitment objective is to minimize expected policy-specific loss, and the second selector minimizes its sample counterpart on the selection set. With an exact follower best response and the specified tie rule, this is a Stackelberg commitment problem. For finite $\beta$, the objective uses the specified logit response.

\textbf{Registered main result.} The main policy-selection comparison used the endogenous catalog, publication-score quality $q=0.70$, 30\% adoption, $\beta=4$, and a one-period listing delay. For each development limit, $K=1$ and $K=\infty$, we chose the policies and three cost levels on 200 selection instances. We fixed those choices and averaged the six cells with equal weight on 200 separate evaluation instances. The gain was $+4.763$ [1.156, 8.345] points, or 0.72\% of the mean status-quo loss of 662.7 points. The paired 95\% interval excluded zero, but its upper endpoint of 1.26\% on the status-quo scale fell below the prespecified 2\% magnitude criterion.

\begin{table*}[!t]
\caption{Loss comparisons at selected costs in the public-record variant. Developed share is the expected share under the status-quo response. Positive differences favor score-first, and positive gains favor probability-only. Brackets give 95\% intervals in loss points~\cite{ref-artifact}.}\label{tab:public-record-shares}
\centering
\ruledtablestyle
\begin{tabularx}{\textwidth}{|c|c|R|R|R|R|}
\hline
& & \multicolumn{2}{c|}{\textbf{Evidence-first minus score-first loss}} & \multicolumn{2}{c|}{\textbf{Probability-only gain over score-first}} \\
\cline{3-4}\cline{5-6}
\textbf{$\kappa$} & \textbf{Developed share} & \multicolumn{1}{c|}{\textbf{Replay}} & \multicolumn{1}{c|}{\textbf{Policy-specific}} & \multicolumn{1}{c|}{\textbf{Replay}} & \multicolumn{1}{c|}{\textbf{Policy-specific}} \\ \hline
0.177 & 0.785 & $+21.1$ [12.9, 29.6] & $+91.8$ [82.8, 101.1] & $-9.4$ [$-23.8$, 4.9] & $-91.3$ [$-107.0$, $-75.5$] \\ \hline
0.354 & 0.660 & $-15.7$ [$-23.9$, $-7.0$] & $+79.3$ [69.3, 89.9] & $+33.3$ [18.4, 47.8] & $-77.5$ [$-95.4$, $-59.8$] \\ \hline
0.500 & 0.545 & $-49.0$ [$-57.2$, $-40.2$] & $+43.2$ [31.9, 55.2] & $+66.2$ [51.0, 81.2] & $-58.6$ [$-79.1$, $-38.0$] \\ \hline
0.707 & 0.384 & $-83.0$ [$-91.1$, $-74.7$] & $-55.3$ [$-69.0$, $-41.1$] & $+98.8$ [83.4, 113.9] & $-29.3$ [$-53.4$, $-5.9$] \\ \hline
1.000 & 0.203 & $-82.1$ [$-88.9$, $-74.8$] & $-181.9$ [$-197.2$, $-165.9$] & $+98.8$ [84.9, 112.6] & $-6.8$ [$-33.8$, 19.1] \\ \hline
1.682 & 0.041 & $-27.7$ [$-31.9$, $-23.7$] & $-138.2$ [$-151.6$, $-124.6$] & $+41.7$ [33.4, 50.3] & $-7.7$ [$-30.5$, 15.2] \\ \hline
2.000 & 0.020 & $-15.1$ [$-18.1$, $-12.3$] & $-90.6$ [$-102.3$, $-79.1$] & $+25.7$ [19.4, 32.6] & $-10.9$ [$-30.7$, 8.6] \\ \hline
\end{tabularx}
\end{table*}

The ranking reversals in Section~\ref{sec-5} use full adoption and concern particular pairs of rules at specified costs. The estimated selection gain concerns a separate policy-selection decision under partial adoption, averaged over six cells.

\textbf{Prespecified supplementary tests.} In the matched six cells at $q=0.55$, the selectors chose the same policies and the gain was exactly zero. At $q=0.70$, the gain had positive point estimates at two adjacent costs in the unlimited-capacity window. Averaged over both development limits and all nine costs, the gain at $q=0.70$ exceeded the gain with an exact publication score by 1.55 [0.36, 2.74] points. The corresponding intervals for $q=0.55$ and $q=0.90$ included zero, leaving the pattern across score qualities uncertain. A decoupled-catalog comparison draws listing calendars from the endogenous status-quo distribution and separates the calendar draw from the executed plan. This keeps the marginal distribution of calendars and removes their dependence on the plan, while the attacker still responds to each policy. By contrast, fixed-response replay keeps the pairing of plans and calendars and uses the same plan probabilities for both rules. The difference in selection gains between the endogenous and decoupled settings was 4.77 [0.94, 8.60] points, below the registered 2\% magnitude criterion.

\textbf{Listing delay.} A prespecified comparison that enters no verdict holds the status-quo policy fixed and delays each listing by two periods. Holding the plans and probabilities of the shorter delay fixed gives the fixed-plan effect, recomputing the response under the longer delay gives the total, and their difference is the response component. Averaged equally across both development limits and all nine costs at $\alpha=0.3$ and $\beta=4$, the increases in status-quo loss were 48.6 [39.4, 58.2] at $q=0.55$ (30.6 with plans held fixed and 17.9 from the changed response), 18.2 [10.6, 26.0] at $q=0.70$ (11.9 and 6.3), $-16.5$ [$-22.4$, $-10.7$] at $q=0.90$ ($-8.9$ and $-7.7$), and $-28.3$ [$-33.4$, $-23.2$] at $q=1.00$ ($-15.3$ and $-13.0$). For this status-quo policy, earlier listing lowered mean loss at $q=0.55$ and $0.70$ and raised it at $q=0.90$ and $1.00$. The sign change agrees with Section~\ref{sec-4}, where score-first has lower loss than evidence-first at low and intermediate costs when the publication score is exact. Until a vulnerability is listed, the status quo's 80\% evidence allocation is unused and goes to the publication-score queue (Table~\ref{tab-1}). Earlier listing draws that budget into the proportional evidence layer sooner. The fixed-plan components show that this lowers loss when the score orders poorly and raises it when the score is accurate.

\textbf{Exploratory configurations.} Among the three exploratory configurations, full adoption with an exact best response was the only one to meet the direction and magnitude criteria. The gain there was 2.38\% of status-quo loss (16.47 [1.49, 31.35] points). This descriptive result suggests a gain under those conditions~\cite{ref-artifact}. The registered main test concerns the partial-adoption configuration above.

\subsection{Public-record attribute shares and sparse development}\label{sec-6-2}

The October 27, 2026 National Vulnerability Database (NVD) snapshot gives 244,134 nonrejected CVEs published from April 14, 2021 to October 26, 2026. The selected CVSS v3 vector was network for 172,112 of 232,217 CVEs (0.74), a proxy for internet-facing exposure. The latest valid Stakeholder-Specific Vulnerability Categorization (SSVC) automatability assessment in NVD was positive for 46,430 of 199,667 CVEs (0.23). By the cohort cutoff, 1,007 (0.41\%) were in KEV, with listing-lag quartiles of 1, 15, and 117 days. The artifact gives the selection rules and source digests~\cite{ref-artifact}. The variant replaces the two default shares (both 0.3) and retains all other synthetic inputs. Table~\ref{tab:public-record-shares} describes 800 fresh evaluation instances, with intervals for probability-only ranking conditional on a predictor fitted to 400 other fresh instances.

In the registered evidence-first comparison, the expected number of developments out of eight falls from about six at $\kappa=0.177$ to 3.2--3.7 at $\kappa=0.595$ and 0.4--0.5 at $\kappa=1.68$~\cite{ref-artifact}. The prespecified reversal tests use costs with relatively dense development. In the public-record variant, replay and policy-specific point estimates had opposite signs in 5 of 21 cost cells for evidence-first versus score-first and in 14 cells for probability-only versus score-first. Both pointwise intervals excluded zero in 3 and 6 of these cells, respectively. In those cells, the expected developed share under the status quo was 38\% to 70\%.

At $\kappa=1.682$ and $2.000$, the developed shares under the status quo were 4.1\% and 2.0\%. In these cells, replay credited probability-only ranking with gains of 41.7 and 25.7 points, with intervals above zero. Policy-specific gains were $-7.7$ and $-10.9$ points, with both intervals crossing zero. Selecting the rule on replay alone would therefore favor a benefit that remains unresolved after adoption. For evidence-first versus score-first, both evaluations favored evidence-first, but the gap was $-27.7$ versus $-138.2$ points at $\kappa=1.682$ and $-15.1$ versus $-90.6$ at $\kappa=2.000$ (Table~\ref{tab:public-record-shares}).

\subsection{Scaling beyond exact enumeration}\label{sec-6-3}

We repeated the evidence-first versus score-first comparison on 60 new seeds at each of two instance sizes while retaining the registered configuration of that comparison. We used exact enumeration at $n=16$ and a Metropolis approximation to the logit response at $n=16$ and $n=32$ (Appendix~\ref{app:plan-sampling}). All estimates in this subsection are descriptive.

Table~\ref{tab:larger-instances} sets the two larger sizes beside the registered comparison at the costs around the registered reversal. At both sizes the replay and policy-specific point estimates have opposite signs at $\kappa=0.595$ and $0.707$. With 32 vulnerabilities at $\kappa=0.707$ both intervals exclude zero, and the gaps are 6.0\% and 8.6\% of status-quo loss in opposite directions, against 3.0\% and 4.7\% in the registered comparison. Both evaluations favor score-first at $\kappa=0.177$ and $0.354$ and evidence-first at $\kappa=1$ and above. Across the seven tested costs, the largest difference between a sampled and an enumerated mean gap at $n=16$ was 1.84 loss points. At $\kappa=1.68$, where 3.6\% of the 32 vulnerabilities were developed under the status-quo response, the replay gap was about one fifth of the policy-specific gap ($-132.6$ against $-641.3$)~\cite{ref-artifact}.

\begin{table}[!t]
\caption{Evidence-first minus score-first loss at three instance sizes, in loss points. Positive values favor score-first. Brackets give 95\% intervals. The first row is the registered comparison on 1,600 instances; the larger sizes use 60 instances each.}\label{tab:larger-instances}
\centering
\ruledtablestyle
\begin{tabularx}{\columnwidth}{|l|c|C|C|}
\hline
\textbf{Instances} & \textbf{$\kappa$} & \textbf{Replay} & \textbf{Policy-specific} \\ \hline
$n=8$, exact & 0.595 & \intervalentry{-23.3}{-28.5,\,-17.8} & \intervalentry{+36.0}{28.2,\,43.9} \\ \hline
$n=16$, exact & 0.595 & \intervalentry{-28.4}{-73.6,\,19.2} & \intervalentry{+139.6}{72.3,\,212.2} \\ \hline
$n=16$, exact & 0.707 & \intervalentry{-63.3}{-107.0,\,-17.8} & \intervalentry{+65.6}{-7.3,\,143.5} \\ \hline
$n=32$, sampled & 0.595 & \intervalentry{-71.8}{-150.3,\,12.8} & \intervalentry{+382.8}{282.7,\,493.7} \\ \hline
$n=32$, sampled & 0.707 & \intervalentry{-163.2}{-241.1,\,-79.8} & \intervalentry{+233.7}{113.1,\,368.7} \\ \hline
\end{tabularx}
\end{table}

The comparison at $n=16$ checks aggregate loss estimates against exact enumeration. At $n=32$, the intervals summarize variation across the 60 instances using sampled losses. They do not quantify error from imperfect mixing, which Appendix~\ref{app:plan-sampling} assesses separately. With chains started from the empty and the full plan, split $\hat R$ of the loss is at most 1.05 for every chain set at six of the seven costs and for 93.3\% of them at $\kappa=1$. The mean gaps differ from the uniform-start estimates by at most 3.1 loss points. In a run four times as long on the first 20 instances, the largest split $\hat R$ is 1.014.

\section{Discussion and limitations}\label{sec-7}

\subsection{What replay measures}\label{sec-7-1}

Time alignment governs which information a patching rule may use at each decision. Fixed-response replay additionally assumes that every candidate rule faces the same distribution of development plans and listing dates. The reversals in Section~\ref{sec-5} show that a ranking can depend on that assumption even when both rules use only information available at the time.

The rule comparisons and the policy-selection experiment answer different questions. The former ask which of two specified rules yields less loss under each evaluation. The latter asks how much a selector gains from modeling response when choosing among budget splits, at 30\% actual adoption. The 43\% crossing in Fig.~\ref{fig:adoption-belief} varies believed adoption at full actual adoption and concerns neither. Within the rule comparisons, evidence-plus-score and the product rule show that conclusions can depend on how a ranking uses each signal.

\subsection{Implications for evaluation}\label{sec-7-implications}

Within the model's assumptions, the results suggest three practices.

\textbf{Match the evaluation to the scale of the decision.} By Proposition~\ref{prop:small-adoption}, an organization with a small adoption share can rank its own losses by replay. Its per-unit loss under replay converges to the loss under policy-specific responses, and the rankings agree when the limiting gap is nonzero. A rule intended for wide adoption, such as a platform default or a directive like BOD 26-04, is a different decision. There the response can change non-adopters' losses at the same order as the adopters' direct effect, and replay alone does not settle the ranking. Our comparisons concern the synthetic budget allocations of Section~\ref{sec-3}. Evaluating the directive itself would require its remediation timelines, agency procedures, and outcomes in federal deployments.

\textbf{Refitting a predictor does not validate the rule that ranks by it.} Pooled and damped refits stabilized the predictor, yet at the intermediate cost the probability-only rule still had higher loss than score-first while replay credited it with a gain. The product rule, which keeps the publication score in the ranking, had a nonnegative gain estimate at all three costs tested.

\textbf{Report loss alongside predictive accuracy.} Concordance with exploitation outcomes and the loss of a rule that ranks by the forecast are separate quantities, and here they diverged once the attacker responded. An evaluation of a forecast as a prioritization tool should state the loss of the rule that uses it and the attacker response it assumes.

\subsection{Limitations and scope}\label{sec-7-3}

Development costs and publication-score quality are synthetic and not calibrated to data, so the experiments cannot place an operational setting on either side of an observed crossing. One descriptive variant replaces two synthetic attribute shares with proportions from public vulnerability records, using a network attack vector as a proxy for internet-facing exposure. The catalog mechanism simplifies observation and reporting (Section~\ref{sec-3-1}), and empirical work would need to distinguish changes in attacker behavior from changes in detection or reporting.

The developed shares in Section~\ref{sec-6-2} are not exploitation rates. FIRST reports that its EPSS data partners observe exploitation activity for about 2.5--3\% of published CVEs in a 30-day window~\cite{ref-3}, whereas a developed share of 2.0\% here is 0.16 expected developments per instance of eight synthetic vulnerabilities over eight model periods. Comparable rates for a remediation queue would need observations restricted to that queue.

The registered comparisons assume that attackers know the rule and its adoption share, and Section~\ref{sec-3-5} treats the believed share as an external input. The attacker's gross return equals the defender's loss, so development cost is the only departure from zero-sum payoffs, and the payoff includes neither diminishing returns nor a motive for stealth. With a finite development limit, the attacker chooses only which vulnerabilities to develop, not their order.

The registered reversals hold for the tested full-adoption settings in Table~\ref{tab-3}, and the capacity, adoption, and delay variants are descriptive. The prediction experiment studies a frozen logistic predictor fitted to simulated exploitation probabilities. Evaluating EPSS itself would require its issued forecasts and the corresponding outcomes.

\section{Conclusion}\label{sec-8}

Fixed-response replay compares patching rules against a fixed history of exploitation and so leaves out how the rules themselves change the attacker's incentives. In our leader--follower game, reallocating a limited patch budget changes how long vulnerabilities stay exposed and can reverse replay's ranking of two rules, as it did for evidence-first against score-first and for probability-only ranking against score-first. Replay and policy-specific evaluation are the two ends of a family indexed by the adoption share the attacker believes. Their gap is bounded through development costs and the rule's effects across plans. For a small adopter, replay ranks its own losses correctly in the limit, although the population-wide ranking can differ. Pooling or damping a predictor's refits curbs their alternation without removing replay's misranking at an intermediate cost, whereas keeping the publication score in the ranking left the rule no worse than score-first, in point estimate, at the costs tested.

These conditions are synthetic. Applying them to an operational setting requires estimates of development costs and score quality, and evidence on how attackers adjust after a widely adopted change in remediation policy.

\section*{Reproducibility}

The repository~\cite{ref-artifact} holds the specifications under \nolinkurl{docs/signal_feedback/}: the model in \nolinkurl{SPEC.md} and the four registered analyses in \nolinkurl{SCORE_QUALITY_SPEC.md}, \nolinkurl{POLICY_BOUNDARY_SPEC.md}, \nolinkurl{RESPONSE_FLIP_SPEC.md}, and \nolinkurl{EXPLOIT_PREDICTION_SPEC.md}. In the same directory, \nolinkurl{ARTIFACT_README.md} maps these files to specification versions, runners, seeds, and output paths. It also indexes the five descriptive supplements, including the sampler checks. The repository includes the source code, tests, result summaries, and checksum lists. A single command reruns the model round and the four registered analyses from their seeds and checks the results against the archived summaries.

\appendices
\section{Proof of the adoption-share mixing identity}\label{app:adoption-scaling}

For Lemma~\ref{lem:adoption-scaling}, fix an instance, a plan $S$ with its development and listing dates, and each rule's queue order. Let $x_t^{r,g}$ be the unpatched counts before period $t$ repairs for a group of share $g\in[0,1]$ using rule $r$. Let $R_{r,t}(x,b)$ return the counts left after repairs with starting counts $x$ and budget $b$. With population counts $N=(N_i)_i$ and budget $B$, the recurrence is
\[
x_0^{r,g}=gN,\qquad x_{t+1}^{r,g}=R_{r,t}(x_t^{r,g},gB).
\]

Baseline repair multiplies published counts by a fixed factor. The proportional tier assigns $y_i\min(\theta b,Y)/Y$ repairs to each listed vulnerability with count $y_i$ when their sum $Y$ is positive, and spends zero otherwise. In a score-ordered queue, each step spends the minimum of the remaining budget and the eligible count. A common scaling of $x$ and $b$ scales repairs, residual counts, and unused budget by the same factor. Thus $R_{r,t}(gx,gb)=gR_{r,t}(x,b)$, also at $g=0$, when both sides vanish.

Induction gives $x_t^{r,g}=g x_t^{r,1}$. The start-of-period loss is $\sum_{i:w_i\le t}s_i x_{t,i}^{r,g}$, so the group incurs $g\ell_r(S)$ over the horizon. Adding the losses of the two groups with shares $a$ and $1-a$ gives $a\ell_\pi(S)+(1-a)\ell_{\pi_0}(S)$.

\section{Sampling development plans}\label{app:plan-sampling}

The larger-instance experiment in Section~\ref{sec-6-3} uses full adoption and $K=\infty$, so each plan $S$ is a subset whose members complete development at publication. For each instance, cost, and response rule $r\in\{\pi_0,\pi_E,\pi_S\}$, the sampler targets the logit distribution proportional to $\exp(\beta u_r(S)/v_{\mathrm{unit}})$, where $u_r(S)=\ell_r(S)-C(S)$.

\begin{enumerate}
\item Initialize 32 chains by including each vulnerability independently with probability $1/2$.
\item For each sweep, draw a uniform random permutation of vulnerability indices, shared across the chains. For each index, propose toggling its membership in each chain's current plan $S$ to obtain $S'$. Accept each proposal with probability
\[
\min\!\left\{1,\exp\!\left(
   \frac{\beta[u_r(S')-u_r(S)]}{v_{\mathrm{unit}}}
\right)\right\}.
\]
Each single-coordinate proposal is symmetric. Its Metropolis update preserves the target distribution, as does the composition in a sweep.
\item Discard 15 sweeps, then retain one plan per chain after each of 30 sweeps, for 960 draws per response distribution. Draws within a chain are correlated.
\end{enumerate}

Replay averages the paired loss differences on the status-quo draws. Policy-specific evaluation averages each rule's loss over its own draws before subtracting the two means. The recorded numerical check compares both mean gaps with exact enumeration at $n=16$.

\textbf{Dispersed-start diagnostics at $n=32$.} On the same 60 instances, we started 16 chains at the empty plan and 16 at the full plan. We computed split $\hat R$ and an autocorrelation-based effective sample size (ESS) for target-rule loss and development count at each cost and target rule. At six of seven costs all 180 chain sets had loss $\hat R\le1.05$ (including the sparse cost $\kappa=1.68$). At $\kappa=1$, 93.3\% were at most 1.05, 98.3\% at most 1.10, and the maximum was 1.229. Median estimated loss ESS reached the 960-draw ceiling at every cost. The fifth percentile ranged from 284 to 960. Development-count diagnostics showed a similar pattern. Mean loss gaps differed from the uniform-start estimates by at most 3.1 points. With 60 burn-in and 120 retained sweeps on the first 20 instances, the maximum $\hat R$ was 1.014 and the smallest fifth-percentile ESS was about 1,559 of 3,840 draws across both quantities and all costs. Mean per-instance wall times for seven costs and three response distributions were 90 and 344 seconds, respectively. These checks measure agreement and autocorrelation for the two recorded summaries~\cite{ref-artifact}.

\end{document}